\documentclass[
aps,
prc,
amsmath,amssymb,
twocolumn,
superscriptaddress,
showkeys,
nofootinbib,
]{revtex4-1}

\usepackage{graphicx}
\usepackage[svgnames]{xcolor}
\usepackage{siunitx}
\usepackage[caption=false]{subfig}
\usepackage[colorlinks=true, citecolor=Blue, linkcolor=Blue, urlcolor=Blue]{hyperref}

\usepackage{etoolbox}
\makeatletter
\appto\abstract{%
  \let\latexlist\list
  \def\list{\edef\keeprightskip{\the\rightskip}\latexlist}%
  \patchcmd\latexlist{\ignorespaces}{\rightskip\keeprightskip\ignorespaces}{}{}%
}
\makeatother

\begin{document}

\title{Collinear cluster tri-partition: 
Kinematics constraints and stability of collinearity
}

\author{P. Holmvall}
 \email[]{holmvall@chalmers.se}
 \affiliation{Department of Physics, Chalmers University of Technology, SE-41296 Gothenburg, Sweden}
 \affiliation{Institut Laue Langevin, 71 avenue des Martyrs, F-38042 Grenoble Cedex 9, France}
\author{U. K\"oster}
 \affiliation{Institut Laue Langevin, 71 avenue des Martyrs, F-38042 Grenoble Cedex 9, France}
\author{A. Heinz}
 \affiliation{Department of Physics, Chalmers University of Technology, SE-41296 Gothenburg, Sweden}
\author{T. Nilsson}
 \affiliation{Department of Physics, Chalmers University of Technology, SE-41296 Gothenburg, Sweden}

\date{\today}

\begin{abstract}
\begin{description}
\item[Background] A new mode of nuclear fission has been proposed by the FOBOS collaboration, called Collinear Cluster Tri-partition (CCT), suggesting that three heavy fission fragments can be emitted perfectly collinearly in low-energy fission. This claim is based on indirect observations via missing-energy events using the $2v2E$ method. This proposed CCT seems to be an extraordinary new aspect of nuclear fission. It is surprising that CCT escaped observation for so long given the relatively high reported yield, of roughly $0.5\%$ relative to binary fission. These claims call for an independent verification with a different experimental technique.

\item[Purpose] Verification experiments based on direct observation of CCT fragments with fission fragment spectrometers require guidance with respect to the allowed kinetic energy range, which we present in this paper. Furthermore, we discuss corresponding model calculations which, if CCT is found in such verification experiments, could indicate how the breakups proceed. Since CCT refers to collinear emission, we also study the intrinsic stability of collinearity.

\item[Methods] Three different decay models are used that together span the timescales of three-body fission. These models are used to calculate the possible kinetic energy ranges of CCT fragments by varying fragment mass splits, excitation energies, neutron multiplicities and scission-point configurations. Calculations are presented for the systems ${^{235}\text{U}}(\text{n}_{\text{th}},\text{f})$ and ${^{252}\text{Cf}}(\text{sf})$, and the fission fragments previously reported for CCT, namely isotopes of the elements Ni, Si, Ca and Sn. In addition, we use semi-classical trajectory calculations with a Monte-Carlo method to study the intrinsic stability of collinearity.

\item[Results] CCT has a high net Q-value, but in a sequential decay, the intermediate steps are energetically and geometrically unfavorable or even forbidden. Moreover, perfect collinearity is extremely unstable, and broken by the slightest perturbation.

\item[Conclusions] According to our results, the central fragment would be very difficult to detect due to its low kinetic energy, raising the question of why other $2v2E$ experiments could not detect a missing-mass signature corresponding to CCT. Considering the high kinetic energies of the outer fragments reported in our study, direct-observation experiments should be able to observe CCT. Furthermore, we find that a realization of CCT would require an unphysical fine-tuning of the initial conditions. Finally, our stability calculations indicate that, due to the pronounced instability of the collinear configuration, a prolate scission configuration does not necessarily lead to collinear emission, nor does equatorial emission necessarily imply an oblate scission configuration. In conclusion, our results enable independent experimental verification and encourage further critical theoretical studies of CCT.
\end{description}
\end{abstract}



\maketitle

\section{\label{sec:introduction}Introduction}
Nuclear fission has been the focus of intense experimental and theoretical studies ever since its discovery almost 80 years ago \cite{bib:hahn39a,bib:meitner39,bib:bohr39}. Usually, fission results in two fragments (binary fission) with similar (symmetric fission) or dissimilar (asymmetric fission) masses. The possibility of fission into three fragments (ternary fission, see \citet{bib:gonnenwein05} for a review), was proposed \cite{bib:present40} shortly after the discovery of binary fission. Experimental evidence of ternary fission was found 70 years ago in nuclear emulsion photographs \cite{bib:tsien46,bib:tsien47a} and in measurements with ionization chambers \cite{bib:farwell47}. Detailed investigations showed that ternary fission occurs once every few hundred fission events. In $90\%$ of all ternary fission events, the third particle, called the ternary particle, is a ${^{4}\text{He}}$ nuclei, and in $9\%$ hydrogen or a heavier helium nuclei. In only $1\%$ of all ternary fission events does the ternary particle have $Z>2$, with yields rapidly dropping with increased $Z$ \cite{bib:gonnenwein04}. Ternary particles up to $Z=16$ have been observed at yields of the order of $10^{-9}$ per fission \cite{bib:tsekhanovich03}. However, early claims \cite{bib:tsien47b,bib:rosen50,bib:muga63,bib:muga67_1,bib:muga67_2} for yet heavier ternary particles or even ``true ternary fission'' with three fragments of comparable masses remain disputed. Dedicated counting experiments searching for such events in planar geometry \cite{bib:schall87} and radiochemical experiments \cite{bib:roy61,bib:stoenner66,bib:kugler71} gave upper yield limits below $10^{-8}$ for true ternary fission.

Therefore, it came as a great surprise when the FOBOS collaboration reported new experiments indicating true ternary fission events with a yield of $5 \cdot 10^{-3}$ per fission \cite{bib:pyatkov10,bib:pyatkov10_3,bib:pyatkov12}. These experiments were performed with the fission fragment spectrometers FOBOS and mini-FOBOS \cite{bib:ortlepp98}, in which detector modules are placed at opposite sides ($180^{\circ}$ angle) of a thin fission target. The fission target backing creates an intrinsic asymmetry of the setup since fragments detected in one of the arms have to traverse the backing. 
Binary coincidences from ${^{252}\text{Cf}(\text{sf})}$ and ${^{235}\text{U}(\text{n}_{\text{th}},\text{f})}$ were measured with this setup. The binary spectrum showed an enhancement of events with lower energy from the detector arm on the side of the target backing. Some of these missing energy events were interpreted as missing mass, that could correspond to a third particle missing detection due to scattering in the fission target backing. The claim was that three heavy fragments were emitted perfectly collinearly along the fission axis, the two lightest fragments in the same direction as the target backing, and the heavy in the opposite, and that the smallest of the three fragments (the ternary particle) did not reach the active area of the detector. Hence, this interpretation was dubbed ``Collinear Cluster Tri-partition'' (CCT). In the following, we will use this definition of CCT as collinear fission events with a relative angle between fragment emission directions of $180\pm2^{\circ}$ \cite{bib:pyatkov10}.

A similar experiment, but without an explicit asymmetry in any of the flight paths, was performed by \citet{bib:kravtsov99}, showing no indication of neither CCT nor missing mass events, down to a level of $7.5\cdot10^{-6}$ per fission in ${^{252}\text{Cf}(\text{sf})}$.

Given these surprising results and the high reported yield of $0.5\%$, the fact that no indication of CCT was found before in neither radiochemical analysis, nor coincidence measurements, calls for an independent verification, preferably with a direct observation method. This is indeed possible, and under way, with the LOHENGRIN fission fragment recoil separator (to be reported in a future paper). For a verification experiment based on direct observation, it is crucial to know which kinetic energies to scan. Since the FOBOS collaboration did not report at which kinetic energies the fragments were measured, these kinetic energies need to be inferred from theory, which is the main focus of this paper. The kinetic energy distribution of one fragment in a ternary decay cannot be derived from first principles. Instead, the full range of kinetic energies allowed by energy and momentum conservation can be calculated, which is done in this study. This is straightforward since CCT is a one-dimensional decay in which the acceleration is repulsion dominated, yielding a limited amount of possibilities of how the kinetic energy can be distributed between the fragments. The possible kinetic energies are reduced even further by the constraint posed by the FOBOS experiments, that two of the fragments have a kinetic energy which is high enough to enter the detector arms and leave a clear signal. An experiment that can cover all the energies allowed by energy and momentum conservation can thus verify CCT model-independently. If events are found, our model calculations would indicate how the breakups proceed in CCT, by comparison with the measured kinetic energies.

We start this paper by detailing which fissioning systems will be studied. This is followed by a description of the theoretical models spanning the possible kinetic energies in CCT, and the Monte-Carlo method used to study the intrinsic stability of collinearity. Results are then presented in the form of possible final kinetic energies in CCT, benchmarks of the methods used, studies that highlight overlooked contradictions in the models currently favored in the literature, and studies of the stability of CCT. This is followed by discussions on verification of CCT by direct observation, on the CCT interpretation and on the intrinsic stability of CCT. Finally, the paper ends with conclusions and appendices with details of each model and method.

\section{\label{sec:systems}Fissioning systems}
In this paper, we present new and detailed calculations on the reported CCT clusters \cite{bib:pyatkov10,bib:pyatkov10_3,bib:pyatkov12}
\begin{eqnarray}
\label{eq:intro:235u_decay}
{^{235}\text{U}}(\text{n}_{\text{th}},\text{f}) & \rightarrow & {\text{Sn}} + {\text{Si}} + {\text{Ni}} + \nu{\cdot}\text{n},
\\
\label{eq:intro:252cf_decay}
{^{252}\text{Cf}}(\text{sf}) & \rightarrow & {\text{Sn}} + {\text{Ca}} + {\text{Ni}} + \nu{\cdot}\text{n},
\end{eqnarray}
both with and without intermediate steps leading up to the final fragments, where $\nu$ is the neutron multiplicity. Other speculated fragments have similar masses and Q-values, and therefore similar kinematics. The derivations presented in this paper allow easy extension to any desired system.

\begin{figure}[b]
\includegraphics[width=1.0\linewidth]{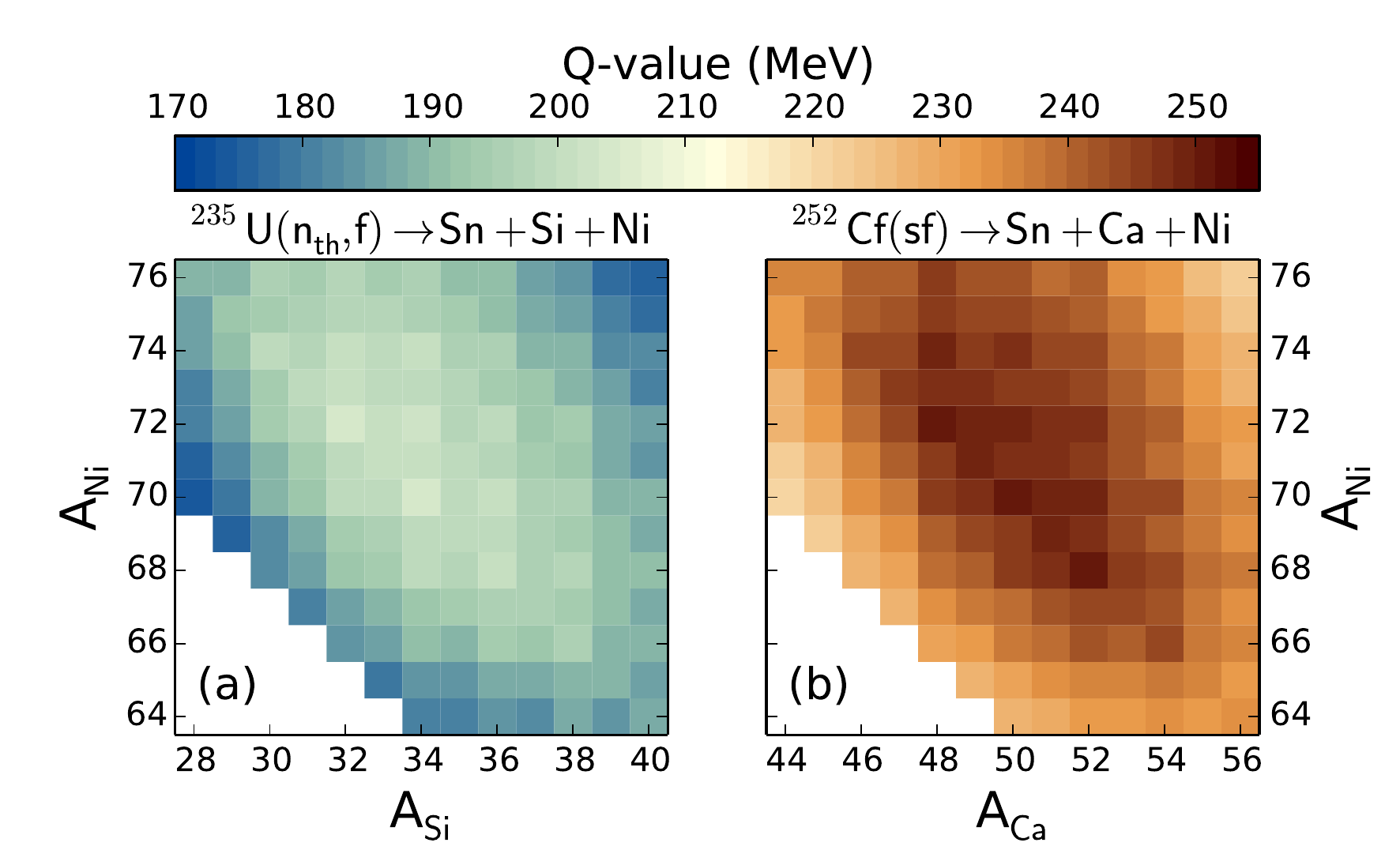}
\caption{\label{fig:fs_qtot} (Color online) Plots (a) and (b) show the Q-value versus the final mass split between the lightest fragments, in the decays in Eqs.~(\ref{eq:intro:235u_decay}) and (\ref{eq:intro:252cf_decay}), respectively, at zero neutron multiplicity ($\nu=0$). The Q-values are calculated from mass excesses taken from AME2012 \cite{bib:audi12}. No data are available for the bottom left corner (i.e. for masses $\text{A}_{\text{Sn}} > 138$). Prompt neutron emission $\nu>0$ generally lowers the Q-values (see Fig.~\ref{fig:fs_q_first}).}
\end{figure}

In the analysis of the FOBOS experiments \cite{bib:pyatkov10,bib:pyatkov10_3,bib:pyatkov12}, the measurements were interpreted as masses $\mathrm{A_{Sn}}\approx132$ and $\mathrm{A_{Ni}}\approx68\text{--}72$ with $\nu \approx 0\text{--}4$, with missing masses $\mathrm{A_{Si}}\approx34\text{--}36$ and $\mathrm{A_{Ca}}\approx48\text{--}52$, respectively. These are the most energetically favorable masses, as shown in Fig.~\ref{fig:fs_qtot}.
Our study includes a slightly wider range of masses. The figure shows Q-values which are relatively high compared to binary fission. As our results will show, however, a high Q-value does not necessarily imply a high yield or probability for fission, since the intermediate steps may be unfavorable or forbidden.

\section{\label{sec:models}Theoretical models}
\begin{figure}[b]
\includegraphics[width=0.65\linewidth]{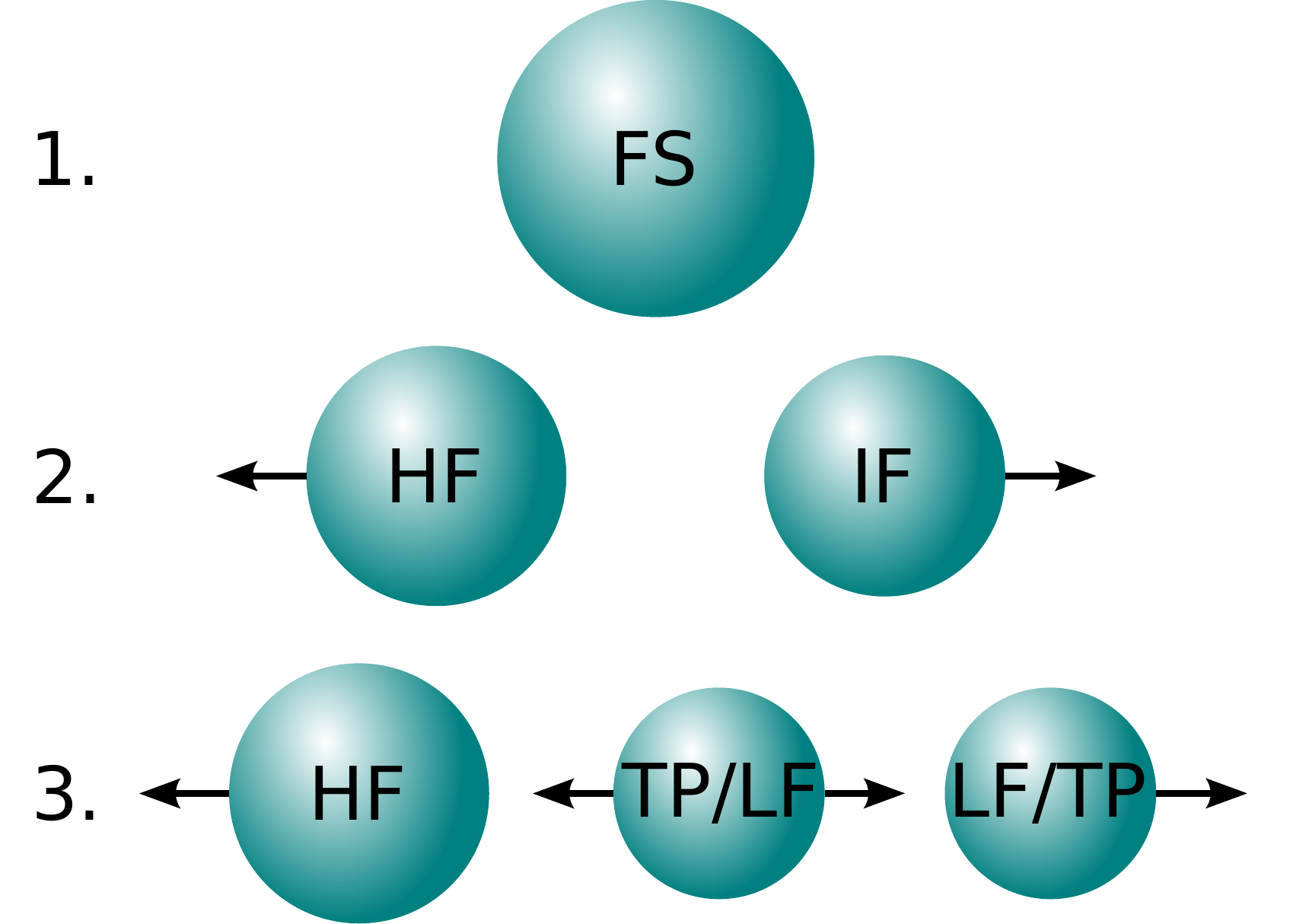}
\caption{\label{fig:sequential_decay} (Color online) Schematic picture of the formation of CCT. For long timescales between two successive (sequential) breakups, there is an intermediate state (2). For sufficiently short timescales between breakups, there is no intermediate state, and the decay is a true three-body decay. Arrows indicate momentum direction. See text for explanation of acronyms. }
\end{figure}

CCT is a decay in one spatial dimension, in which three fragments are formed from one fissioning system (FS) through two breakups \cite{bib:garrido06, bib:kadmensky16} and accelerated along the same line (see Fig.~\ref{fig:sequential_decay}). If the time between breakups is long enough, there exists an intermediate state with a heavy fragment (HF) and an intermediate fragment (IF), the latter which splits in turn into a light fragment (LF) and a ternary particle (TP). The ternary particle here refers to the lightest fragment. If the time between breakups is sufficiently short, there is no intermediate state, and the decay is a ``true'' three-body decay.
Therefore, for a given fissioning system, the essential parameter to describe CCT is the time between the two breakups. In this paper, we divide the timescale of this parameter into three regimes, or models, and explicitly show that the kinetic energies of these models overlap in the limits.

The first model is called the ``sequential'' decay model \cite{bib:vijayaraghavan12} and is based on two sequential binary fissions, with long timescales between the two successive scissions (i.e. assuming fully accelerated fragments before the second scission). The second model is the recently proposed ``almost sequential'' decay model \cite{bib:tashkhodjaev15}, with intermediate timescales between scissions (assuming partially accelerated fragments before the second scission). These sequential models are the currently favored models in the literature. As is shown in the results, however, both of these models assume the fission of an intermediate system with a high fission barrier and extremely low (or even negative) Q-value. This motivates the study of a third model which is based on traditional ternary fission models, and called in the following the ``true ternary'' decay model, with ``infinitesimal'' timescales between scissions (i.e. no intermediate step or fragment). We mainly focus on the sequential and true ternary decay models, as they represent the extremes of the kinetic energy range, but we also show how to calculate all the possible kinetic energies allowed by energy and momentum conservation for all three models, for other fissioning systems as well.

The final kinetic energies of the fission fragments are obtained analytically in the sequential decay model. This is possible since the kinematics in this model is fully determined by energy and momentum conservation. In the almost sequential and true ternary decay models, the final kinetic energies depend on the dynamics. Thus, these results are obtained with semi-classical trajectory calculations (see \citet{bib:wagemans91} chapter 12-III for a review). In these calculations, the scission configuration (initial fragment positions and momenta) is constrained by energy and momentum conservation for a given fissioning system. Subsequently, the final kinetic energies are calculated by starting from the scission configuration and solving the equations of motion iteratively. The latter is done with a fourth-order Runge-Kutta method using a time step of $10^{-25}$~s, until more than $99\%$ of the potential energy is converted into kinetic energy.

Since the 1960s, semi-classical trajectory calculations have been applied to ternary fission, mainly with the aim to determine the scission configuration \cite{bib:boneh67}. As in many of these studies, a ``point charge approximation'' is used in our trajectory calculations, which assumes only a repulsive Coulomb force between spherical fragments. For the purpose of finding which scission configuration matches a particular final distribution, this method has received critique due to ambiguity \cite{bib:dakowski79,bib:gavron75,bib:pikpichak84,bib:kunhikrishnan12}, since several initial configurations can have the same final distribution. We do not have the same aim, however. Instead, we vary all possible initial collinear configurations in order to find all possible final kinetic energies of CCT fragments. Furthermore, we again stress the fact that in contrast to the previously mentioned studies, we study CCT which is a one-dimensional problem in which the dynamics during the fission fragment acceleration is dominated by the repulsive Coulomb interaction. Adding an attractive nuclear correction to the sequential model does not affect the final momenta, since the latter is uniquely determined by energy and momentum conservation. This is verified by the perfect agreement between our results and that of \citet{bib:vijayaraghavan12}, who included an attractive nuclear correction. Still, we show explicitly that the attractive nuclear correction has a negligible effect in both the sequential and the almost sequential decay models (see Sec.~\ref{sec:almost_sequential_decay_results}). In the true ternary decay model, the attractive nuclear interaction reduces the possible kinetic energy range (as discussed in Sec.~\ref{sec:simultaneous_decay_kinematics}). Since we are looking for the widest possible kinetic energy range to cover experimentally, the attractive nuclear interaction is excluded in this model to get a safe upper limit.

In addition to deriving the possible final kinetic energies, we use a Monte-Carlo method to sample perturbations in the trajectory calculations, testing the intrinsic stability of collinearity in ternary fission, yielding the final angular distributions versus the perturbations. Previous studies using the point charge approximation with a Monte-Carlo approach successfully reproduced experimental ternary fission data \cite{bib:guet80,bib:radi82}. Furthermore, for the purpose of calculating final kinetic energies and angular distributions, it has been shown that the simple point charge approximation gives similar results to more sophisticated models, which incorporate attractive nuclear forces, fragment deformations and other effects \cite{bib:carjan80,bib:flassig83}.

Nevertheless, to test the validity of our semi-classical trajectory calculations, we set up several benchmarks. As a quantitative verification against analytical calculations, the sequential and almost sequential models are compared for extremely long times between the two scissions, and the two techniques show excellent agreement (see results in Sec.~\ref{sec:sequential_decay_results}). Additional tests for ternary fission with ${^{4}\text{He}}$ (not reported here) reproduced well the results of the previously mentioned studies. We also verified for certain configurations that the inclusion of higher order moments corresponding to deformed fragments does not considerably affect the final momenta along the fission axis.

\subsection{\label{sec:sequential_decay_kinematics}Sequential decay model}
\begin{figure}[b]
\includegraphics[width=1.0\linewidth]{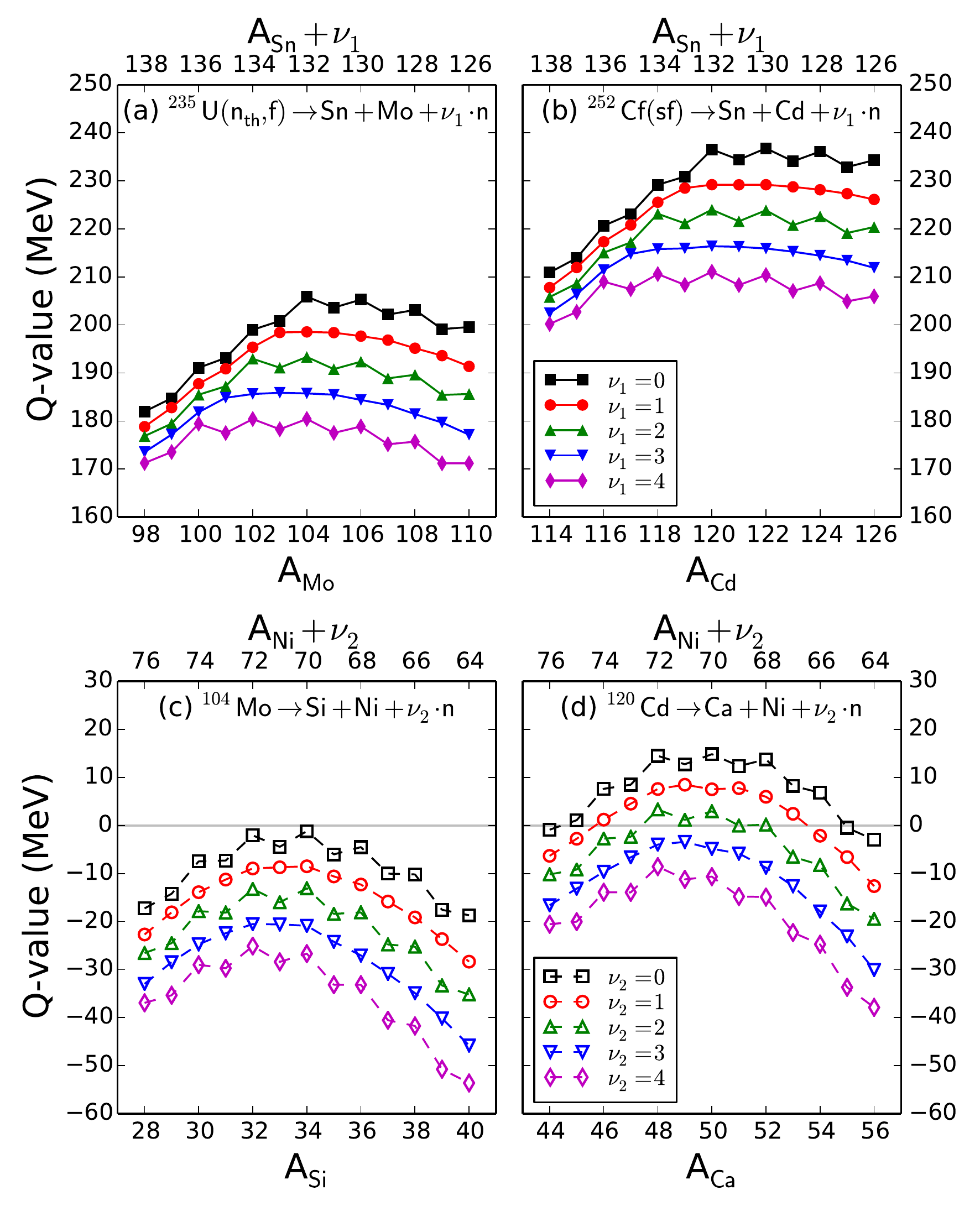}
\caption{\label{fig:fs_q_first} (Color online) The Q-value versus the mass split in the binary decays of (a) ${^{235}\text{U}}(\text{n}_{\text{th}},\text{f})$, (b) ${^{252}\text{Cf}}(\text{sf})$, (c) ${^{104}\text{Mo}}$ and (d) ${^{120}\text{Cd}}$. The Q-values are calculated from the mass excesses, taken from AME2012 \cite{bib:audi12}. Lines have been added to guide the eye.}
\end{figure}

In the sequential decay model \cite{bib:vijayaraghavan12}, the fissioning system splits into a heavy fragment and an intermediate fragment. The latter splits in turn into a light fragment and a ternary particle. The ternary particle here refers to the lightest fragment. Either the TP or the LF can be formed at the center, as illustrated in Fig.~\ref{fig:sequential_decay}. Potential energy surface calculations \cite{bib:vijayaraghavan15,bib:oertzen15,bib:tashkhodjaev15} predict that it is more likely that the TP is formed at the center. Nevertheless, we present results for both cases. Using conservation of proton numbers in Eqs.~(\ref{eq:intro:235u_decay}) and (\ref{eq:intro:252cf_decay}), the intermediate fragments are found to be molybdenum (Mo) and cadmium (Cd) in ${^{235}\text{U}}(\text{n}_{\text{th}},\text{f})$ and ${^{252}\text{Cf}}(\text{sf})$, respectively. Allowing for neutron emission from the FS and the IF with multiplicities $\nu_1$ and $\nu_2$, respectively, gives
\begin{eqnarray}
{^{235}\text{U}}(\text{n}_{\text{th}},\text{f}) & \rightarrow & \text{Sn} + \text{Mo} + \nu_1\cdot\text{n}\nonumber\\
\label{eq:sequential_fission_235u}
& \rightarrow & \text{Sn} + \text{Si} + \text{Ni} + (\nu_1+\nu_2)\cdot\text{n}
\\
{^{252}\text{Cf}}(\text{sf}) & \rightarrow & \text{Sn} + \text{Cd} + \nu_1\cdot\text{n}\nonumber\\
\label{eq:sequential_fission_252cf}
& \rightarrow & \text{Sn} + \text{Ca} + \text{Ni} +  (\nu_1+\nu_2)\cdot\text{n}.
\end{eqnarray}
The most energetically favorable masses of the IFs are found to be $\text{A}_{\text{Mo}}=104$ and $\text{A}_{\text{Cd}}=120$ with neutron multiplicity $\nu_1=0$, as seen in Figs.~\ref{fig:fs_q_first} (a) and (b), respectively. The most favorable mass split in the decay of ${^{104}\text{Mo}}$ is $\text{A}_{\text{Ni}}=70$ and $\text{A}_{\text{Si}}=34$ with $\nu_2=0$, and in the decay of ${^{120}\text{Cd}}$ it is $\text{A}_{\text{Ni}} = 70$ and $\text{A}_{\text{Ca}}=50$ with $\nu_2=0$, as seen in Figs.~\ref{fig:fs_q_first} (c) and (d), respectively.
We will present final kinetic energies for a range of masses centered around these mass splits, with neutron multiplicities $\nu_1=0\text{--}4$. Note, however, that in the decay of both Mo and Cd, the Q-value is extremely low, for many mass splits even negative, and that any neutron multiplicity $\nu_2>0$ lowers the Q-value further. To have any chance of decaying, the IF needs excitation energy (from here on denoted ${E^{*}_{IF}}$). Even if the low Q-values are compensated for by an extremely high excitation energy, it does not mean that the intermediate fragment can fission, it also has to overcome a very high fission barrier (see Sec.~\ref{sec:cct_model_discussion} for discussion). Therefore, we assume cold compact fission of the IF, by setting both the neutron multiplicity $\nu_2$ and the sum of the excitation energies of the final fragments $\mathit{TXE} = {E^{*}_{HF}}+{E^{*}_{TP}}+{E^{*}_{LF}}$ to zero in our calculations. We show how to calculate a more general case, however, and such results can be directly obtained from our results by simple subtraction. Any $\mathit{TXE} > 0$ lowers the sum of the final kinetic energies accordingly, and any $\nu_2>0$ lowers the IF Q-value and the final total kinetic energy of the TP and LF by up to $8$~MeV per neutron (see discussion in Sec.~\ref{sec:cct_model_discussion}).

The final kinetic energies of the fragments will be calculated and presented versus fragment mass splits, neutron multiplicity and the excitation energy ${E^{*}_{IF}}$.

Details of this model are found in App.~\ref{sec:app:almost_sequential}.

\subsection{\label{sec:almost_sequential_decay_kinematics}``Almost sequential'' decay model}
\begin{figure}[b]
\includegraphics[width=0.65\linewidth]{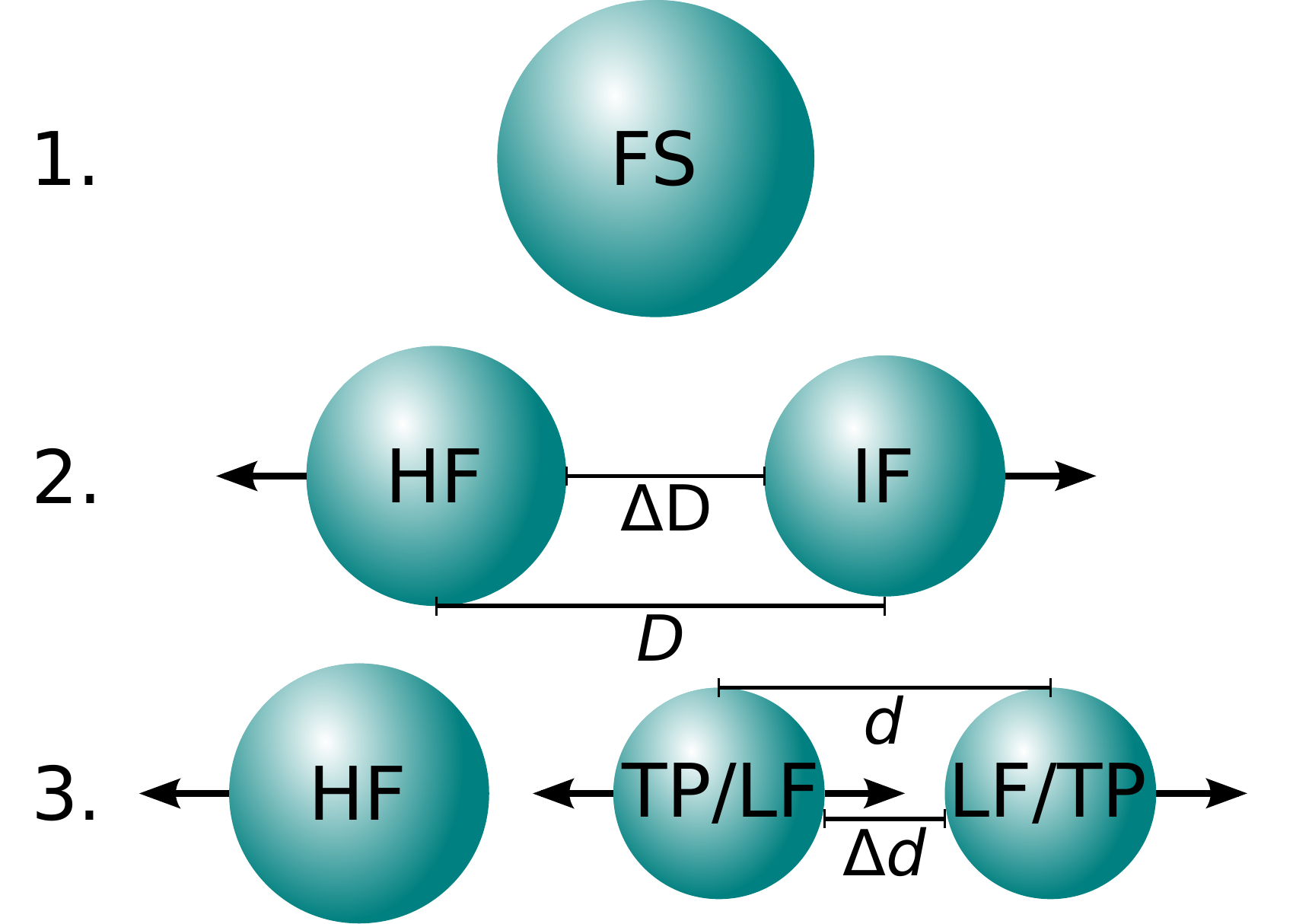}
\caption{\label{fig:almost_sequential_decay} (Color online) CCT as an ``almost sequential'' decay. In contrast to the sequential model, the Coulomb repulsion of the heavy fragment is crucial after the second scission, making the inter-fragment distances relevant to the kinematics. Arrows indicate momentum direction.}
\end{figure}

To calculate the kinematics of an ``almost sequential'' decay \cite{bib:tashkhodjaev15}, a similar parametrization as in the sequential model is used. The main difference with respect to the sequential model is the finite time between the first and the second scission, which is analogous to the charge-center distance between the HF and the IF at the second scission, denoted $D$ (see Fig.~\ref{fig:almost_sequential_decay}).
The finite time and distance makes it necessary to account for the Coulomb repulsion at all stages in the almost sequential model, thus the final kinetic energies depend on the full dynamics. To this end the scission-point configuration after the second scission is constrained, and the final kinetic energies are calculated from this configuration using semi-classical trajectory calculations (described in the beginning of this section). As will be shown in the results (Sec.~\ref{sec:almost_sequential_decay_results}), an attractive nuclear interaction is found to have negligible influence on the final kinetic energies.

Apart from the parameters of the sequential model (neutron multiplicities, fragment mass splits and excitation energies), the almost sequential model relies on two additional parameters to constrain the scission-point configuration. We choose these parameters to be the
tip distances (surface separation distances) between the HF and the IF ($\Delta{D}$) at the moment of the second scission, and between the LF and the TP ($\Delta{d}$) after the second scission. The tip distance is defined as
\begin{equation}
\label{eq:seq:asec:tip_distance}
\Delta{D}_{ij} = D_{ij} - R_{i} - R_{j},
\end{equation}
where $R_{k}=r_0\sqrt[3]{A_k}$ is the radius of fragment $k$ with mass $A_k$ and $r_0\approx1.25$~fm, and $D_{ij}$ is the charge-center distance between the respective fragments. Note that as $D,\Delta{D} \to \infty$, the equations of the almost sequential decay model become exactly the same as those for the sequential decay model. Details of this model are found in App.~\ref{sec:app:almost_sequential}.

As will be shown in the results (Sec.~\ref{sec:almost_sequential_decay_results}), not even the most favorable mass splits will have enough energy to allow for a physically reasonable tip distance ($<4$~fm \cite{bib:gonnenwein91}). Therefore, cold compact fission of the IF is assumed in our calculations, i.e. minimizing $\Delta{d}$, by setting both the neutron multiplicity $\nu_2$ and the sum of the final fragment excitation energies $\mathit{TXE}=E^{*}_{HF}+E^{*}_{TP}+E^{*}_{LF}$ to zero. As described in Sec.~\ref{sec:sequential_decay_kinematics}, results for $\nu_2>0$ and $\mathit{TXE}>0$ can be obtained directly from our results.

\subsection{\label{sec:simultaneous_decay_kinematics}True ternary decay model}
\begin{figure}[b]
\includegraphics[width=0.65\linewidth]{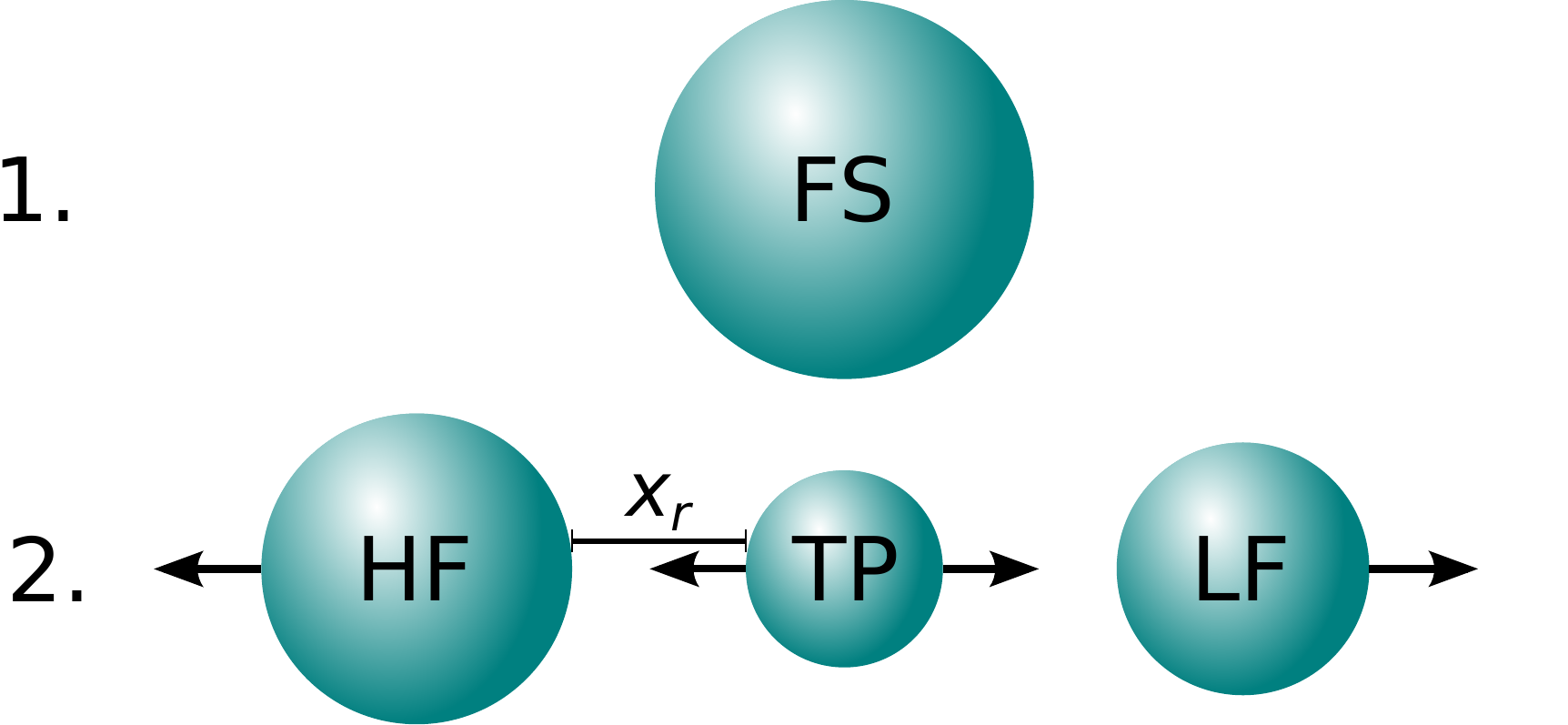}
\caption{\label{fig:simultaneous_decay} (Color online) CCT as a true ternary decay. Arrows indicate momentum direction.}
\end{figure}

In the most common theoretical models of ternary fission (see \citet{bib:wagemans91} chapter 12 and references therein), all three fragments are considered to be formed during a very short time interval from the same fissioning system, with the ternary particle at the center. The different models have a similar parametrization, but are based on different hypotheses and favor different starting positions of the ternary particle between the heavier fragments. Our true ternary decay model is based on the most common models, but is collinear, as is illustrated in Fig.~\ref{fig:simultaneous_decay}.
Furthermore, our model treats the ternary particle offset between the other fragments as a parameter, denoted $x_{r}$. We let $x_{r} = 0$ and $x_{r} = 1$ correspond to the cases when the ternary particle is formed ``touching'' the heavy and light fragment, respectively. The results show that the highest kinetic energy for the LF is achieved if the TP is formed touching the HF. This is because the HF accelerates the TP, which then transfers momentum to the LF. The opposite configuration gives the lowest kinetic energy for the LF, and the highest possible for the HF. Obviously, these touching configurations are not real ``scission'' configurations, since in reality the fragments would not separate due to the attractive nuclear force. Therefore, the exploration from one touching configuration to the other will predict a wider kinetic energy range than physically possible. The touching configurations thus provide safe upper limits for the experimental search, which is why the attractive nuclear interaction is disregarded in this model.

The scission-point configuration is constrained by energy conservation for given fragment mass splits, neutron multiplicity and pre-scission kinetic energies, with the parameters $x_r$ and $\mathit{TXE}$, where the latter is the sum of the fragment excitation energies ($\mathit{TXE} = E^{*}_{HF} + E^{*}_{TP} + E^{*}_{LF}$). Note that axial pre-scission kinetic energy can be canceled in most cases by choosing an earlier reference time, corresponding to a tighter scission configuration. We have therefore set the pre-scission kinetic energy to zero in our calculations. Lateral pre-scission kinetic energy is studied in Sec.~\ref{sec:intrinsic_stability_collinearity}, and is found to break collinearity, even for extremely low values.

Using the scission-point configuration, the final kinetic energies are computed with semi-classical trajectory calculations, as described in Sec.~\ref{sec:models}.

Details of this model are found in App.~\ref{sec:app:ternary}.

\subsection{\label{sec:intrinsic_stability_collinearity}Intrinsic stability of collinearity}
\begin{figure}[b]
\includegraphics[width=0.65\linewidth]{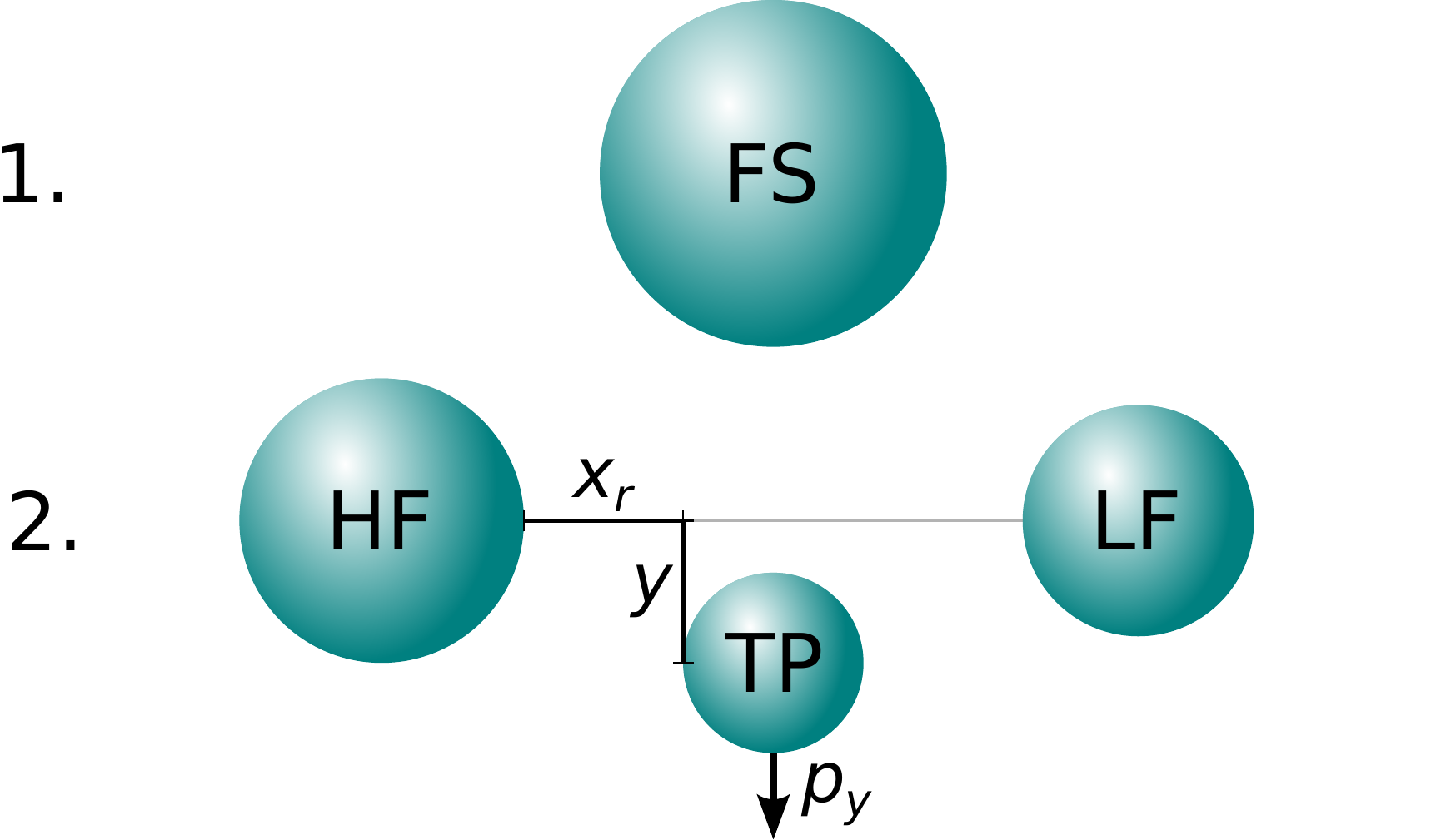}
\caption{\label{fig:stability_simultaneous_decay} (Color online)
CCT as a true ternary decay, with an initial lateral momentum or spatial offset of the ternary particle from the fission axis. The arrow indicates momentum direction.}
\end{figure}

Using the true ternary decay model (Sec.~\ref{sec:simultaneous_decay_kinematics}), the intrinsic stability of collinearity in ternary fission is analyzed by using a Monte-Carlo method to sample a perturbation in the ternary particle position and momentum perpendicular to the fission axis, independently (see Fig.~\ref{fig:stability_simultaneous_decay}).
As in the true ternary decay model, the parameters are $x_r$ (the relative ternary particle position at scission, as described in Sec.~\ref{sec:simultaneous_decay_kinematics}) and $\mathit{TXE}$ (the sum of the fragment excitation energies). In addition, a parameter representing the perturbation is also varied, being either initial lateral momentum or spatial offset of the ternary particle from the fission axis, denoted $p_y$ and $y$, respectively. Given these parameters, the scission-point configuration is uniquely constrained by invoking conservation of energy, as well as linear and angular momentum. Each parameter is sampled in a uniform interval, with $\sim100$ sampling points per parameter, giving more than $10^{6}$ data points per system.

Using the scission-point configuration, the final kinetic energies and emission angles are computed with semi-classical trajectory calculations, as described in Sec.~\ref{sec:models}.

Details of this model are found in App.~\ref{sec:app:stability}.

\section{\label{sec:results}Results}
The results are divided into four subsections. The first subsection covers final kinetic energies in the sequential decay model, both if the ternary particle is formed at the center (which is considered the most favorable case according to potential energy surface calculations \cite{bib:vijayaraghavan15,bib:oertzen15,bib:tashkhodjaev15}), and if the light fragment is formed at the center, for sake of completeness. The first subsection also includes a benchmark of the semi-classical trajectory calculations, which is used in the other models.

The second subsection covers results for the ``almost sequential'' decay model, which show that this model spans the kinetic energy continuum between the sequential and ``true ternary'' decay models. Furthermore, it is explicitly shown that although CCT might have a high net Q-value, the intermediate steps in a sequential and an almost sequential decay are energetically and geometrically unfavorable or even forbidden. It is also shown that the attractive nuclear interaction is negligible in both the sequential models.

The third subsection covers final kinetic energies in the true ternary decay model.

The fourth subsection covers an analysis of the intrinsic stability of collinearity in ternary fission, in which the final scattering angle between the ternary particle and light fragment is presented versus a spatial and a momentum-based perturbation, independently. Requiring a collinear emission sets a threshold on the position and momentum of the ternary particle, which is shown to be much smaller than variations expected due to the uncertainty principle.

\subsection{\label{sec:sequential_decay_results}Sequential decay results}
\begin{figure}[t]
%
\includegraphics[width=1.0\linewidth]{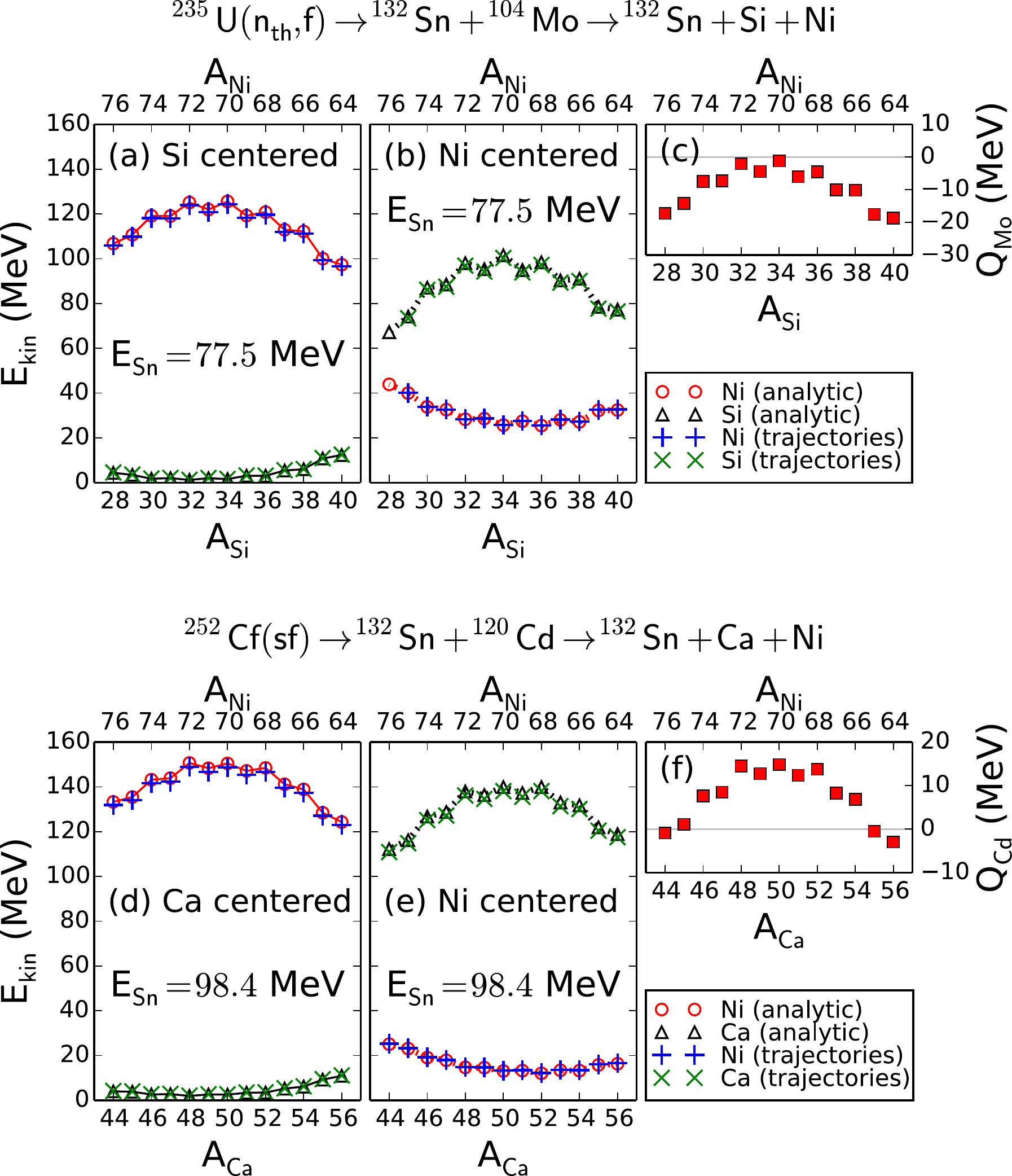}
\caption{\label{fig:sequential_double} (Color online) Final kinetic energies of the TP and LF versus the TP to LF mass split in the sequential decay of (a)--(b) ${^{235}\text{U}}(\text{n}_{\text{th}},\text{f})$ and (d)--(e) ${^{252}\text{Cf}}(\text{sf})$, calculated with the analytic method ($\bigcirc$, $\bigtriangleup$) described in Sec.~\ref{sec:sequential_decay_kinematics} (derivations in App.~\ref{sec:app:almost_sequential}), and with trajectory calculations (+, $\times$) described in Sec.~\ref{sec:almost_sequential_decay_kinematics} (derivations in App.~\ref{sec:app:almost_sequential}). The case when the TP is formed at the center is shown in (a) and (d), while the case when the LF is formed at the center is shown in (b) and (e), (upper and lower signs in Eqs.~(\ref{eq:aseq:p_lf}) and (\ref{eq:aseq:p_tp}), respectively). The corresponding final kinetic energy of the HF is labeled in each plot. The excitation energy of the intermediate fragment is $E^{*}_{IF} = 30$~MeV. The missing trajectory calculations for $A_{\text{Si}}=28$ highlights that the intermediate steps of the decay are energetically forbidden. The corresponding Q-values in the fission of (c) ${^{104}\text{Mo}}$ and (f) ${^{120}\text{Cd}}$ are calculated from mass excesses taken from AME2012 \cite{bib:audi12}. Lines have been added to guide the eye.
}
\end{figure}

Using the sequential model described in Sec.~\ref{sec:sequential_decay_kinematics} (derivations in App.~\ref{sec:app:almost_sequential}), we present in Figs.~\ref{fig:sequential_double} (a)--(b) and (d)--(e) the final fragment kinetic energies versus the mass split between the TP and the LF in the decays
\begin{eqnarray}
\label{eq:seq:235u}
{^{235}\text{U}}(\text{n}_{\text{th}},\text{f}) & \rightarrow & {^{132}\text{Sn}} + {^{104}\text{Mo}} \rightarrow {^{132}\text{Sn}} + \text{Si} + \text{Ni} \\
\label{eq:seq:252cf}
{^{252}\text{Cf}}(\text{sf}) & \rightarrow & {^{132}\text{Sn}} + {^{120}\text{Cd}} \rightarrow {^{132}\text{Sn}} + \text{Ca} + \text{Ni},
\end{eqnarray}
respectively (note that the mass split between the HF and the IF will be varied later). For sake of completeness, results are presented for fission both when the (a),(d) TP and when (b),(e) the LF are formed at the center. Figs.~\ref{fig:sequential_double} (c) and (f) show the Q-value in the fission of the intermediate fragments ${^{104}\text{Mo}}$ and ${^{120}\text{Cd}}$, respectively.
In both systems, the excitation energy of the intermediate fragment is $E^{*}_{IF} = 30$~MeV. As a benchmark of the semi-classical trajectory calculations, the figures also show results (+, $\times$) obtained from the almost sequential model (described in Sec.~\ref{sec:almost_sequential_decay_kinematics}, derivations in App.~\ref{sec:app:almost_sequential}) at extremely long times between the two scissions. There is an excellent agreement between the two methods, as shown by the complete overlap of the symbols. The small difference is attributed to the fact that the trajectory calculations have to start and end with a finite potential energy ($<1\%$). The kinetic energy of the HF is labeled in each plot. The shape of the kinetic energy plot directly follows the Q-value in the fission of the intermediate fragment.

\begin{figure}[b]
\includegraphics[width=1.0\linewidth]{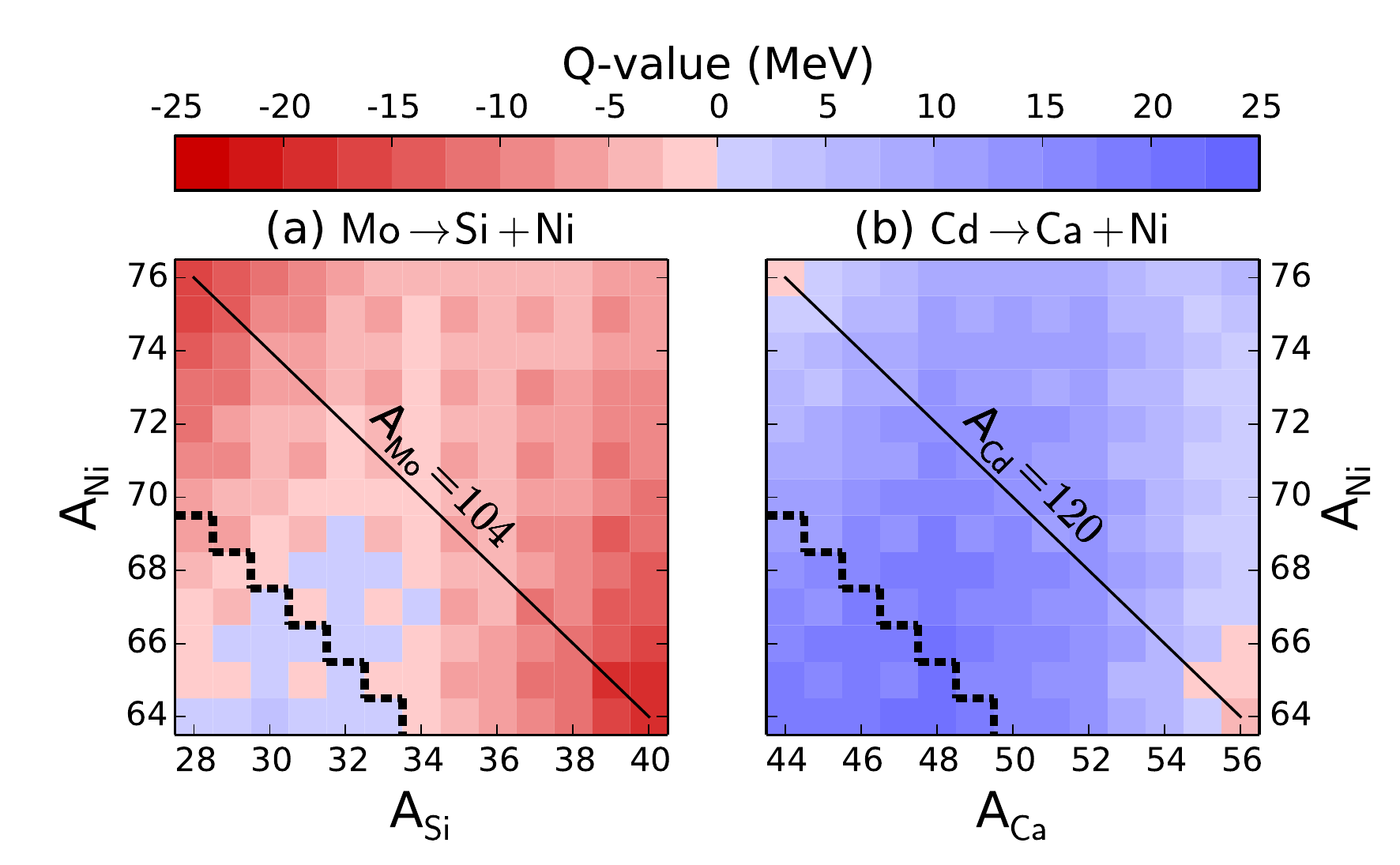}
\caption{\label{fig:IF_q_values} (Color online) The Q-value as a function of the mass split between the LF and the TP, in the decays (a) $\text{Mo} \rightarrow \text{Si} + \text{Ni}$ and (b) $\text{Cd} \rightarrow \text{Ca} + \text{Ni}$. Recall from Fig.~\ref{fig:fs_qtot} that masses under the dashed line correspond to nuclides without data for the corresponding HF ($\text{A}_{\text{Sn}} > 138$), and from Fig.~\ref{fig:fs_q_first} that the most favorable mass splits are $\text{A}_{\text{Sn}}=132$ with (a) $\text{A}_{\text{Mo}}=104$ and (b) $\text{A}_{\text{Cd}}=120$. The most favorable mass split between the TP and LF are therefore found along the solid diagonal lines as $A_{\text{Ni}}=70$ with (a) $A_{\text{Si}}=34$ and (b) $A_{\text{Ca}}=50$. The Q-values are calculated from mass excesses taken from AME2012 \cite{bib:audi12}. Prompt neutron emission from the IF ($\nu_2$) lowers the Q-values significantly (see Fig.~\ref{fig:fs_q_first}).
}
\end{figure}

For comparison, the Q-value in the fission of both Mo and Cd is shown as a function of the mass split between the TP and LF in Fig.~\ref{fig:IF_q_values}.
To have any probability of fissioning, only the most energetically favorable systems should be considered. Further calculations therefore assume that no neutrons originate from the fission of the IF (i.e. $\nu_2=0$ in Eqs.~(\ref{eq:sequential_fission_235u}) and (\ref{eq:sequential_fission_252cf})), and that $\mathit{TXE} = E^{*}_{HF} + E^{*}_{TP} + E^{*}_{LF} = 0$~MeV. Any $\mathit{TXE}> 0$~MeV lowers the final kinetic energy sum accordingly. If any neutrons are emitted in the fission of the IF, the Q-value, and therefore the summed kinetic energy of the TP and LF, are reduced by up to $8$~MeV per neutron. See Sec.~\ref{sec:cct_model_discussion} for further discussion.

\begin{figure*}[tb]
  \includegraphics[width=0.8\linewidth]{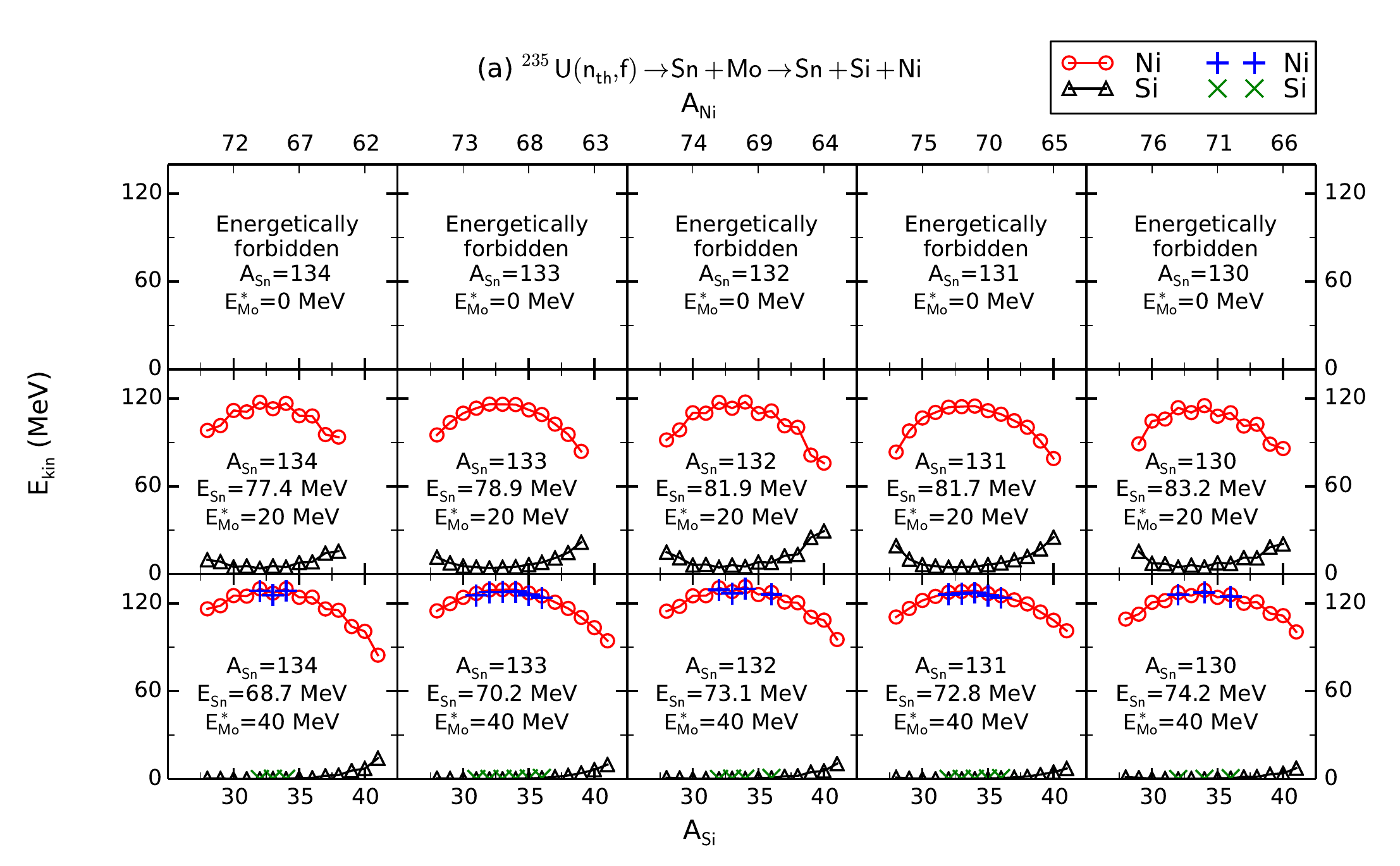}
\hfill
  \includegraphics[width=0.8\linewidth]{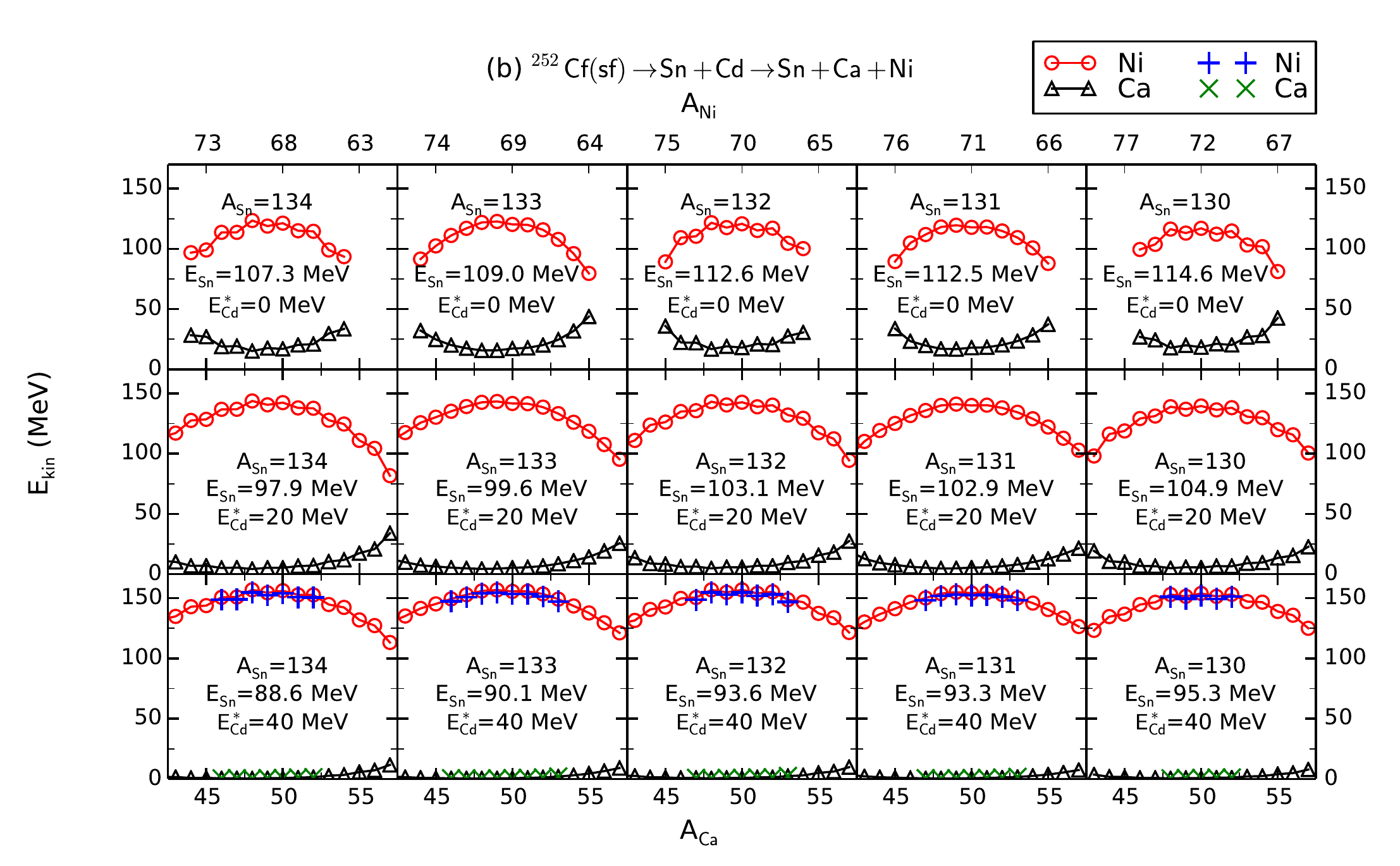}
\caption{\label{fig:sequential_vary_a1} (Color online) Final kinetic energies of the LF and the TP versus their mass split in the sequential decay of (a) ${^{235}\text{U}}(\text{n}_{\text{th}},\text{f})$, and (b) ${^{252}\text{Cf}}(\text{sf})$, when the TP is formed at the center. The kinetic energies are calculated with the analytic method ($\bigcirc$, $\bigtriangleup$) described in Sec.~\ref{sec:sequential_decay_kinematics} (derivations in App.~\ref{sec:app:almost_sequential}), and with trajectory calculations (+, $\times$) described in Sec.~\ref{sec:almost_sequential_decay_kinematics} (derivations in App.~\ref{sec:app:almost_sequential}). Comparing plots in the horizontal direction, the mass split between the HF and the IF is varied, and comparing plots in the vertical direction, $E^{*}_{IF}$ is varied. The values of these parameters and the final kinetic energy of the HF are labeled in each plot. Trajectory calculations are only shown for systems that are energetically allowed and have a tip distance at the second scission of $\leq 7$~fm.
}
\end{figure*}

To see how the kinetic energies are affected when varying $E^{*}_{IF}$ and the mass split between the HF and the IF, multiple plots are compared to each other in a grid in Fig.~\ref{fig:sequential_vary_a1}, for both (a) ${^{235}\text{U}}(\text{n}_{\text{th}},\text{f})$, and (b) ${^{252}\text{Cf}}(\text{sf})$. Comparing plots in the horizontal direction, the heavy fragment mass is varied $A_{\text{Sn}}=134\text{--}130$, and comparing plots in the vertical direction, $E^{*}_{IF}$ is varied ($0$, $20$ and $40$~MeV).
The corresponding final kinetic energy of Sn and the varied parameters are labeled in each plot. Note that the results for ${^{252}\text{Cf}}(\text{sf})$ in Fig.~\ref{fig:sequential_vary_a1} (b) shows perfect agreement with the corresponding parameter choices in Fig.~6 of \citet{bib:vijayaraghavan12}. Increasing the excitation energy $E^{*}_{IF}$ frees more energy for the acceleration of the TP and LF. Because the direction of acceleration of the inner fragment is opposite to the flight direction of the IF before it fissions, the inner fragment is retarded. For higher excitation energies, the kinetic energy of the inner fragment approaches $0$~MeV. Increasing $E^{*}_{IF}$ leaves less energy available as kinetic energy to the HF. Note that some of the corresponding systems are energetically forbidden for lower excitation energies and for unusual $N/Z$ ratios. Results obtained with trajectory calculations (+, $\times$) only show fission decays which are energetically allowed and have a tip distance that is $\leq 7$~fm at the second scission. None of the systems have a tip distance which is considered to be physically valid, i.e. less than $4$~fm \cite{bib:gonnenwein91} (see Sec.~\ref{sec:almost_sequential_decay_results} for results and discussion).

\begin{figure}[b]
\includegraphics[width=1.0\linewidth]{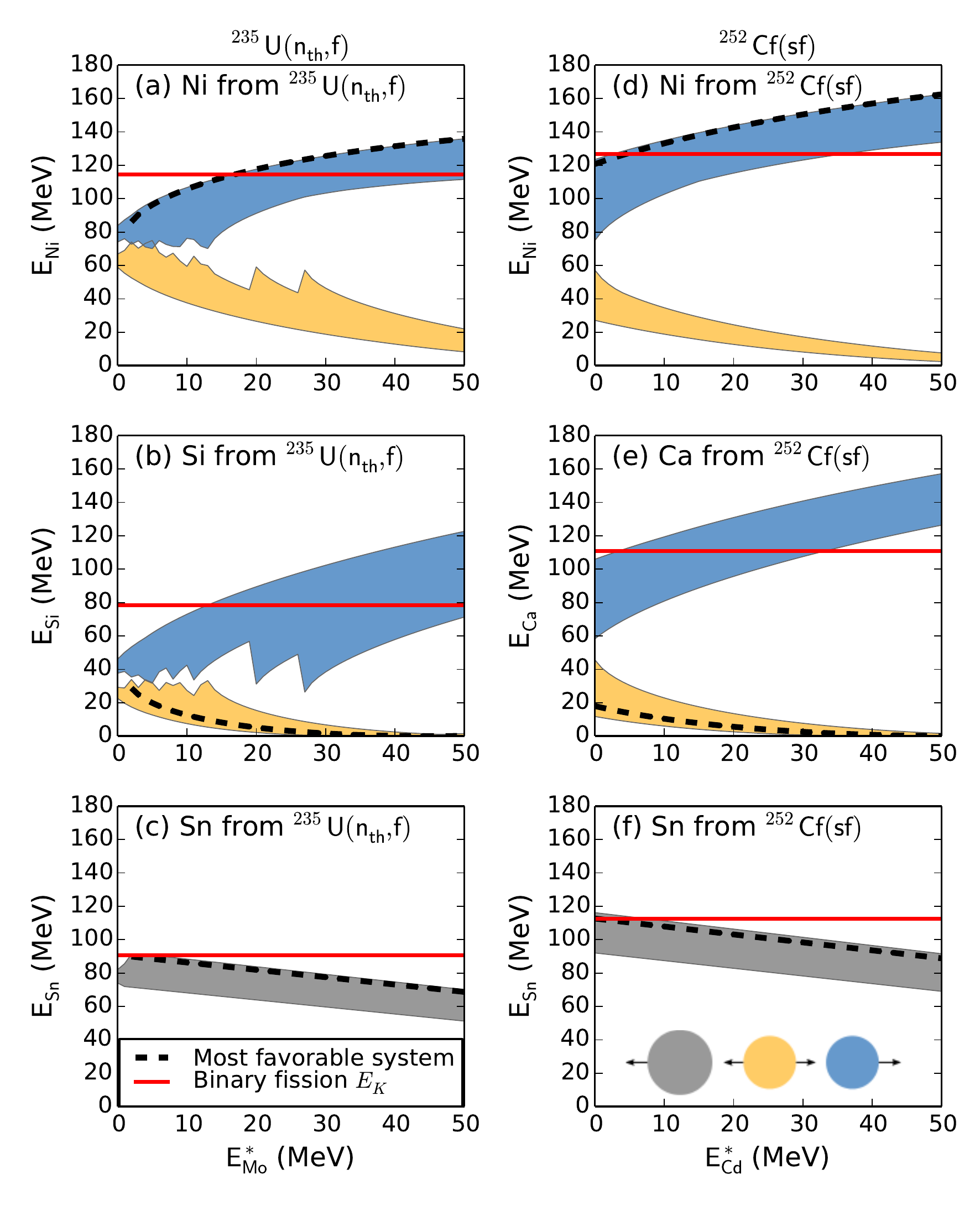}
\caption{\label{fig:sequential_areas_all} (Color online) Areas of attainable final fragment kinetic energies versus the excitation energy of the intermediate fragment in the sequential decays (a--c) ${^{235}\text{U}}(\text{n}_{\text{th}},\text{f}) \rightarrow \text{Sn} + \text{Mo} + \nu_1\cdot\text{n} \rightarrow \text{Sn} + \text{Si} + \text{Ni} + \nu_1\cdot\text{n}$, and (d--f) ${^{252}\text{Cf}}(\text{sf}) \rightarrow \text{Sn} + \text{Cd} + \nu_1\cdot\text{n} \rightarrow \text{Sn} + \text{Ca} + \text{Ni} + \nu_1\cdot\text{n}$. Each figure is element specific, and shows results both when the TP and when the LF are formed at the center. The colors indicate formation position, as shown by the inset in (f). The areas are spanned between the highest and lowest kinetic energies obtained. For a given set of parameters, the final kinetic energy versus ${E^{*}_{IF}}$ follows a well-defined line. The bold dashed line is an example of the latter, with the most favorable set of parameters. For comparison, the horizontal lines correspond to the kinetic energy of the same fragment from compact binary fission (see text and Eqs.~(\ref{eq:u_binary_ni})--(\ref{eq:cf_binary_sn})). The masses are varied as $\text{A}_{\text{Sn}}=128\text{--}134$ and $\text{A}_{\text{Ni}}=68,70,72$, and the prompt neutron multiplicity as $\nu_1=0\text{--}4$. As a consequence, the other masses are in the ranges (a--c) $\text{A}_{\text{Mo}}=98\text{--}108$, $\text{A}_{\text{Si}}=28\text{--}40$, and (d--f) $\text{A}_{\text{Cd}}=114\text{--}124$, $\text{A}_{\text{Ca}}=42\text{--}56$.
}
\end{figure}

Finally, we present in Fig.~\ref{fig:sequential_areas_all} the final kinetic energies when varying all parameters (including the neutron multiplicity) simultaneously, in the sequential decays of (a--c) ${^{235}\text{U}}(\text{n}_{\text{th}},\text{f})$ and (d--f) ${^{252}\text{Cf}}(\text{sf})$, respectively. The parameter ranges are given in the caption.
We represent the kinetic energy range for each fragment with an area, spanned by the highest and lowest kinetic energies obtained. The colors of the areas represent formation configuration (either the TP or the LF is formed at the center), as shown in the inset in Fig.~\ref{fig:sequential_areas_all} (f). The bold dashed lines correspond to the most energetically favorable cluster combinations for each fissioning system, namely
\begin{eqnarray}
{^{235}\text{U}}(\text{n}_{\text{th}},\text{f}) & \rightarrow & {^{132}\text{Sn}} + {^{104}\text{Mo}}\nonumber\\
\label{eq:res:most_favorable_235u}
& \rightarrow & {^{132}\text{Sn}} + {^{34}\text{Si}} + {^{70}\text{Ni}}\\
{^{252}\text{Cf}}(\text{sf}) & \rightarrow & {^{132}\text{Sn}} + {^{120}\text{Cd}}\nonumber\\
\label{eq:res:most_favorable_252cf}
& \rightarrow & {^{132}\text{Sn}} + {^{50}\text{Ca}} + {^{70}\text{Ni}}.
\end{eqnarray}
For a given set of parameters, the kinetic energy of the outer and inner fragment versus $E^{*}_{IF}$ follows an increasing and decreasing curve, respectively. The heavy fragment is not affected by the second scission, and its kinetic energy is therefore linearly decreasing with increasing $E^{*}_{IF}$. The curves are cut off when ${Q_{IF}^{\text{eff}}} < 0$ (as defined in Eq.~(\ref{eq:aseq:q2_eff}) in App.~\ref{sec:app:almost_sequential}), which is why the artificial ``teeth'' structures appear. The horizontal lines correspond to the maximum kinetic energies of the same fragment that would be produced in cold compact binary fission (zero excitation energy and consequently no neutron evaporation) as calculated from Q-values:
\begin{eqnarray}
\label{eq:u_binary_ni}
&\text{(a)}\quad& {^{235}\text{U}(\text{n}_{\text{th}},\text{f})} \rightarrow {^{70}\text{Ni}} + {^{166}\text{Gd}}\\
\label{eq:u_binary_si}
&\text{(b)}\quad& {^{235}\text{U}(\text{n}_{\text{th}},\text{f})} \rightarrow {^{34}\text{Si}} + {^{202}\text{Pt}}\\
\label{eq:u_binary_sn}
&\text{(c)}\quad& {^{235}\text{U}(\text{n}_{\text{th}},\text{f})} \rightarrow {^{132}\text{Sn}} + {^{104}\text{Mo}}\\
\label{eq:cf_binary_ni}
&\text{(d)}\quad& {^{252}\text{Cf}(\text{sf})} \rightarrow {^{70}\text{Ni}} + {^{182}\text{Yb}}\\
\label{eq:cf_binary_ca}
&\text{(e)}\quad& {^{252}\text{Cf}(\text{sf})} \rightarrow {^{50}\text{Ca}} + {^{202}\text{Pt}}\\
\label{eq:cf_binary_sn}
&\text{(f)}\quad& {^{252}\text{Cf}(\text{sf})} \rightarrow {^{132}\text{Sn}} + {^{120}\text{Cd}}.
\end{eqnarray}
The mean kinetic energy of binary fragments lies much lower than these horizontal lines due to the considerable excitation energies of binary fragments. Experiments searching for ternary fission, which are not based on coincidence measurements, can thus use these limits as a reference, in order to determine the source of possible events. If events are found above the maximum energy of binary fission, the origin must be ternary fission.

\subsection{\label{sec:almost_sequential_decay_results}Almost sequential decay results}
It is explicitly shown in the following that since the almost sequential decay model represents the time continuum between the sequential and true ternary decay models, it also represents the kinetic energy continuum. It is also shown that both the sequential and the almost sequential models are geometrically and energetically unfavorable or forbidden, and that the attractive nuclear interaction has a negligible influence on the kinetic energies. The almost sequential model is described in Sec.~\ref{sec:almost_sequential_decay_kinematics} (derivations in App.~\ref{sec:app:almost_sequential}).

\begin{figure}[b]
\includegraphics[width=1.0\linewidth]{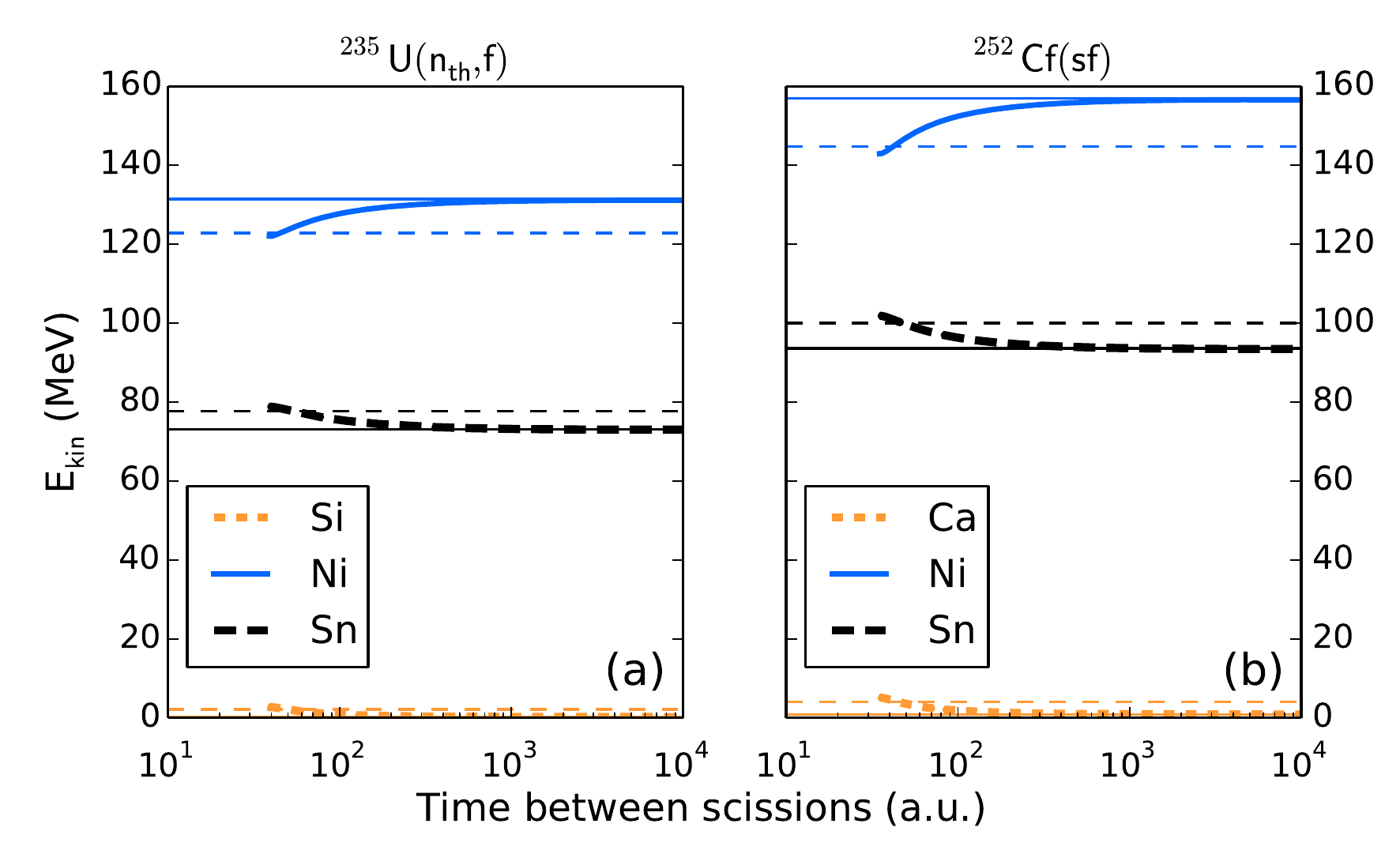}
\caption{\label{fig:almost_sequential_d0_sweep} (Color online) Final kinetic energies versus the time between the first and the second scission (logarithmic scale), in the ``almost sequential'' decays (a) ${^{235}\text{U}}(\text{n}_{\text{th}},\text{f}) \rightarrow {^{132}\text{Sn}} + {^{104}\text{Mo}} \rightarrow {^{132}\text{Sn}} + {^{34}\text{Si}} + {^{70}\text{Ni}}$, and (b) ${^{252}\text{Cf}}(\text{sf}) \rightarrow {^{132}\text{Sn}} + {^{120}\text{Cd}} \rightarrow {^{132}\text{Sn}} + {^{50}\text{Ca}} + {^{70}\text{Ni}}$. The excitation energies are $E^{*}_{IF}=40$~MeV and $\mathit{TXE}=0$~MeV, the tip distance at the second scission is $\Delta{d_0}=7$~fm, and the ternary particle is set to be formed at the center. The fine solid and dashed lines represent the corresponding kinetic energies of the sequential and true ternary decay models, respectively. See text for further explanation how these results are obtained. Less than $1\%$ of the total energy remains as potential energy in the almost sequential and ternary model results.}
\end{figure}

In Fig.~\ref{fig:almost_sequential_d0_sweep}, the final fragment kinetic energies are shown versus the time between the two scissions (i.e. the distance $D$ between the HF and the LF at the second scission, due to the direct correspondence), in the almost sequential decays of (a) ${^{235}\text{U}}(\text{n}_{\text{th}},\text{f})$, and (b) ${^{252}\text{Cf}}(\text{sf})$, respectively.
See caption for mass splits and other parameter values. The results show that as the time between scissions becomes very long, the kinetic energies approach the asymptotic results of the sequential model (fine solid lines), since in the limit $D\to\infty$, the equations of the two models become identical. Furthermore, the results show that at short times between scissions, the kinetic energies approach the asymptotic results of the true ternary decay model (fine dashed lines).
The true ternary decay model results are obtained as described in Sec.~\ref{sec:simultaneous_decay_kinematics} and App.~\ref{sec:app:ternary}, by using the same mass splits, setting $\mathit{TXE}=0$~MeV and using the corresponding distances (setting $D-x = r_{TP}+r_{LF}+\Delta{d}$, which gives $x_r=0.03$ and $x_r=0.02$ for ${^{235}\text{U}}(\text{n}_{\text{th}},\text{f})$ and ${^{252}\text{Cf}}(\text{sf})$, respectively).

We will now derive the maximum $E_{IF}^{*}$ versus the first scission tip distance between the HF and the IF, $\Delta{D_0}$. Furthermore, we will study if this ${E^{*}_{IF}}$ can balance the extremely low Q-values of the IF, for different values of the second scission tip distance between the TP and the LF, $\Delta{d_0}$ (see Fig.~\ref{fig:almost_sequential_decay} for an illustration of these distances). This study will reveal which constraints are posed on the scission configurations if requiring an energetically allowed decay. We focus on the most energetically favorable case, i.e. when the TP is formed at the center, when there is no pre-scission kinetic energy, and with the mass splits from Eqs.~(\ref{eq:res:most_favorable_235u}) and (\ref{eq:res:most_favorable_252cf}) for ${^{235}\text{U}}(\text{n}_{\text{th}},\text{f})$ and ${^{252}\text{Cf}}(\text{sf})$, respectively. We will also assume fully accelerated fragments before the second scission, since the Coulomb barrier in the fission of the IF is higher when the heavy fragment is present. Energy balance of the first fission gives
\begin{equation}
\label{eq:res:first_q}
Q_{FS} = V_1(\Delta{D}) + E_{HF} + E_{LF} + E_{HF}^{*} + E_{IF}^{*},
\end{equation}
where $\Delta{D}$ is the tip distance and $V_1$ the potential
\begin{equation}
\label{eq:res:potential}
V_1 = E_C + E_N.
\end{equation}
Here, $E_C$ is the repulsive Coulomb potential and $E_N$ the attractive nuclear potential. For the latter, we used the Yukawa-plus-exponential function \cite{bib:krappe79_prl,bib:krappe79_prc}. The total excitation energy $\mathit{TXE}_{1} = {E_{IF}^{*}} + {E_{HF}^{*}}$ is maximal when it takes up all the available energy, with zero pre-scission kinetic energy $E_{HF} = E_{LF} = 0$. Thus, the maximum excitation energy for a given scission tip distance $\Delta{D_0}$ is
\begin{equation}
\label{eq:res:TXE_HFIF}
\mathit{TXE}_{1}^{\text{max}} = Q_{FS} - V_1(\Delta{D_0}).
\end{equation}
If this quantity is less than zero, the corresponding scission configuration is energetically forbidden. Consequently, $\mathit{TXE}_{1} = 0$ would give the tightest possible scission configuration (cold compact fission). Energy balance at the second scission gives
\begin{multline}
\label{eq:res:second_q}
Q_{IF} + E_{IF}^{*} + E_{IF} + V_1(D_1) = V_2(\Delta{d_0})\\+ E_{TP} + E_{LF} + E_{TP}^{*} + E_{LF}^{*},
\end{multline}
where $D_1$ is the distance between the HF and the IF (i.e. between the HF and the mass center of the TP and the LF) at the moment of the second scission, and $\Delta{d_0}$ is the TP to LF scission tip distance. Assuming that there is no additional pre-scission kinetic energy of the TP and the LF, other than that from the IF, imposes the constraint $v_{IF} = v_{TP} = v_{LF}$. Assuming also conservation of mass ($m_{IF} = m_{TP} + m_{LF}$), leads to $E_{IF} = E_{TP} + E_{LF}$ right after the second scission. Consequently, for a given scission tip distance $\Delta{d_0}$, the available energy in the second fission becomes
\begin{equation}
\label{eq:res:TXE_TPLF}
E_{TP}^{*} + E_{LF}^{*} = Q_{IF} + E_{IF}^{*} + V_1({D_1}) - V_2({\Delta{d_0}}).
\end{equation}
Again, if this quantity is less than zero, the corresponding scission configuration is energetically forbidden. Assuming fully accelerated fragments before the second scission sets $V_1(D_1\to\infty)\to0$. The maximum available energy or the tightest scission configuration in the second fission can therefore be obtained as a function of $\Delta{d_0}$ and $\Delta{D_0}$, where the latter gives the available $E_{IF}^{*}$. Combining Eqs.~(\ref{eq:res:first_q})--(\ref{eq:res:TXE_TPLF}) gives
\begin{equation}
\label{eq:res:aseq_final}
\mathit{TXE}_{2} = Q_{IF} + \mathit{TXE}_{1} + V_1({D_1}) - V_2({\Delta{d_0}})
\end{equation}
where $\mathit{TXE}_{2} = E_{HF}^{*} + E_{TP}^{*} + E_{LF}^{*}$.
This quantity reflects the net total energy available after the second fission, having provided the corresponding $E_{IF}^{*}$.

\begin{figure}[b]
\centering
\includegraphics[width=1.0\linewidth]{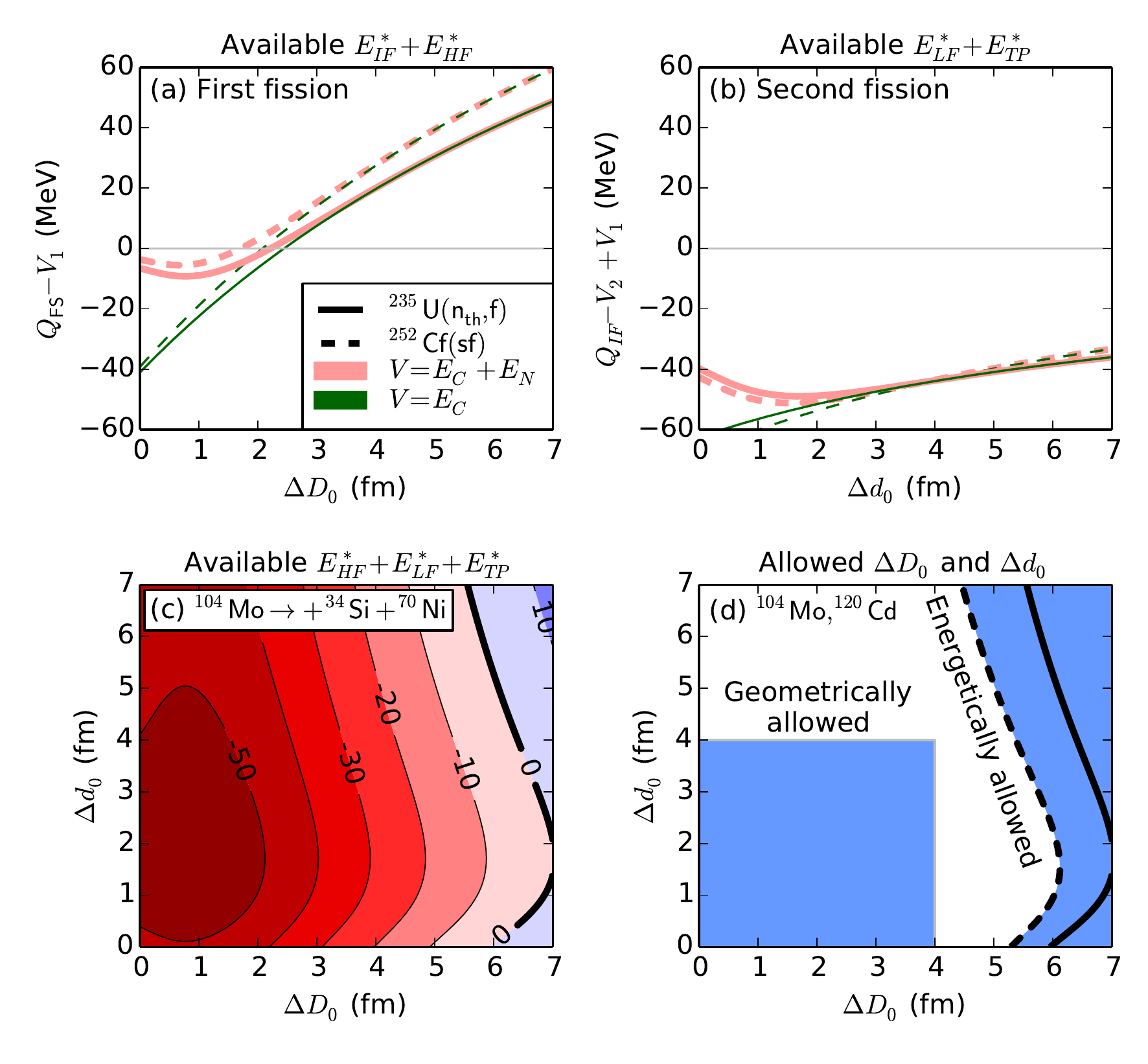}
\caption{\label{fig:solutions_almost_sequential} (Color online) The maximal available energy versus the corresponding scission tip distance in the fission processes (a) ${^{235}\text{U}}(\text{n}_{\text{th}},\text{f}) \rightarrow {^{132}\text{Sn}} + {^{104}\text{Mo}}$ (solid lines) and ${^{252}\text{Cf}}(\text{sf}) \rightarrow {^{132}\text{Sn}} + {^{120}\text{Cd}}$ (dashed lines), as well as in (b) ${^{104}\text{Mo}} \rightarrow {^{34}\text{Si}} + {^{70}\text{Ni}}$ (solid lines) and ${^{120}\text{Cd}} \rightarrow {^{50}\text{Ca}} + {^{70}\text{Ni}}$ (dashed lines). Results are shown both with and without an attractive nuclear interaction (bold and fine lines, respectively). In (c), a contour plot of the maximal available energy after fission of the IF (Eq.~(\ref{eq:res:aseq_final})) for a certain scission tip distance $\Delta{d_0}$ is shown, assuming the maximum $E_{IF}^{*}$ has been provided. The latter is set by the tip distance $\Delta{D_0}$ of the first fission. The first and the second fissioning systems are ${^{235}\text{U}}(\text{n}_{\text{th}},\text{f})$ and ${^{104}\text{Mo}}$, respectively, and the contour lines show the available energy in MeV. In (d), the shaded regions show the $\Delta{D_0}$ and $\Delta{d_0}$ which lead to geometrically and energetically allowed fission of the IF, both for ${^{235}\text{U}}(\text{n}_{\text{th}},\text{f})$ (solid lines) and ${^{252}\text{Cf}}(\text{sf})$ (dashed lines). See text for further information. }
\end{figure}

In Figs.~\ref{fig:solutions_almost_sequential} (a) and (b), the energy available in the first and second fissions are shown versus $\Delta{D_0}$ and $\Delta{d_0}$ (Eqs.~(\ref{eq:res:TXE_HFIF}) and (\ref{eq:res:TXE_TPLF})), respectively. Results are shown for the systems ${^{235}\text{U}}(\text{n}_{\text{th}},\text{f})$ (solid lines) and ${^{252}\text{Cf}}(\text{sf})$ (dashed lines), both with and without an attractive nuclear interaction (bold and fine lines, respectively).
It is apparent that there is a conflict in trying to reduce the first and the second scission tip distances. The reason is that a more narrow first tip distance will leave less $E_{IF}^{*}$ available, but a more narrow second tip distance requires a higher $E_{IF}^{*}$. Keeping both as narrow as possible, an $E_{IF}^{*}$ in the range of $39\text{--}40$~MeV is required in both ${^{235}\text{U}}(\text{n}_{\text{th}},\text{f})$ and ${^{252}\text{Cf}}(\text{sf})$. The figures also show that at any energetically allowed tip distances, the attractive nuclear interaction is negligible, and that it therefore is safe to neglect it in any further analysis. Fig.~\ref{fig:solutions_almost_sequential} (c) shows the total net energy available after the second fission (Eq.~(\ref{eq:res:aseq_final})) as a contour plot versus $\Delta{d_0}$ and $\Delta{D_0}$, where the latter gives the available $E_{IF}^{*}$. The system is ${^{235}\text{U}}(\text{n}_{\text{th}},\text{f})$, and it is inherently assumed that the maximal $E_{IF}^{*}$ has been used. The available energies are indicated along the contour lines in units of MeV. Negative energies indicate that the corresponding scission configurations are energetically forbidden, and the contour line of $0$~MeV shows the most compact scission configuration that is energetically allowed. In Fig.~\ref{fig:solutions_almost_sequential} (d), the values of $\Delta{D_0}$ and $\Delta{d_0}$ that leads to energetically allowed fission of the IF and geometrically allowed scission configurations are indicated by the corresponding shaded areas. With the latter, we refer to that typical tip distances at scission are close to $\sim 2.5$~fm, while tip distances over $4$~fm are generally not considered as physically valid \cite{bib:gonnenwein91}. The solid and dashed lines indicate where ${^{235}\text{U}}(\text{n}_{\text{th}},\text{f})$ and ${^{252}\text{Cf}}(\text{sf})$ are energetically allowed, respectively. Note that for no scission configurations is the decay geometrically and energetically allowed simultaneously. We remind that these results were obtained for the most favorable systems and choice of parameters. Any non-zero excitation energies $E_{HF}^{*}, E_{TP}^{*}, E_{LF}^{*}$ separates the regions in Fig.~\ref{fig:solutions_almost_sequential} (d) even further as indicated by the contour lines, as does any less favorable mass splits, emitted neutrons, pre-scission kinetic energies or finite $D_1$. Even if there was somehow any configurations that were energetically and geometrically allowed, the low fissility and fission barrier penetrability has to be accounted for as well (see discussion in Sec.~\ref{sec:cct_model_discussion}).

In conclusion, these results show that CCT as a sequential decay is geometrically and energetically unfavorable or forbidden. This is highlighted by the fact that there is a competition in keeping the first and the second scission compact, and that a high $E_{IF}^{*}$ is required but leads to less geometrically favorable scission configurations.

\subsection{\label{sec:simultaneous_decay_results}True ternary decay results}
\begin{figure}[b]
\includegraphics[width=1.0\linewidth]{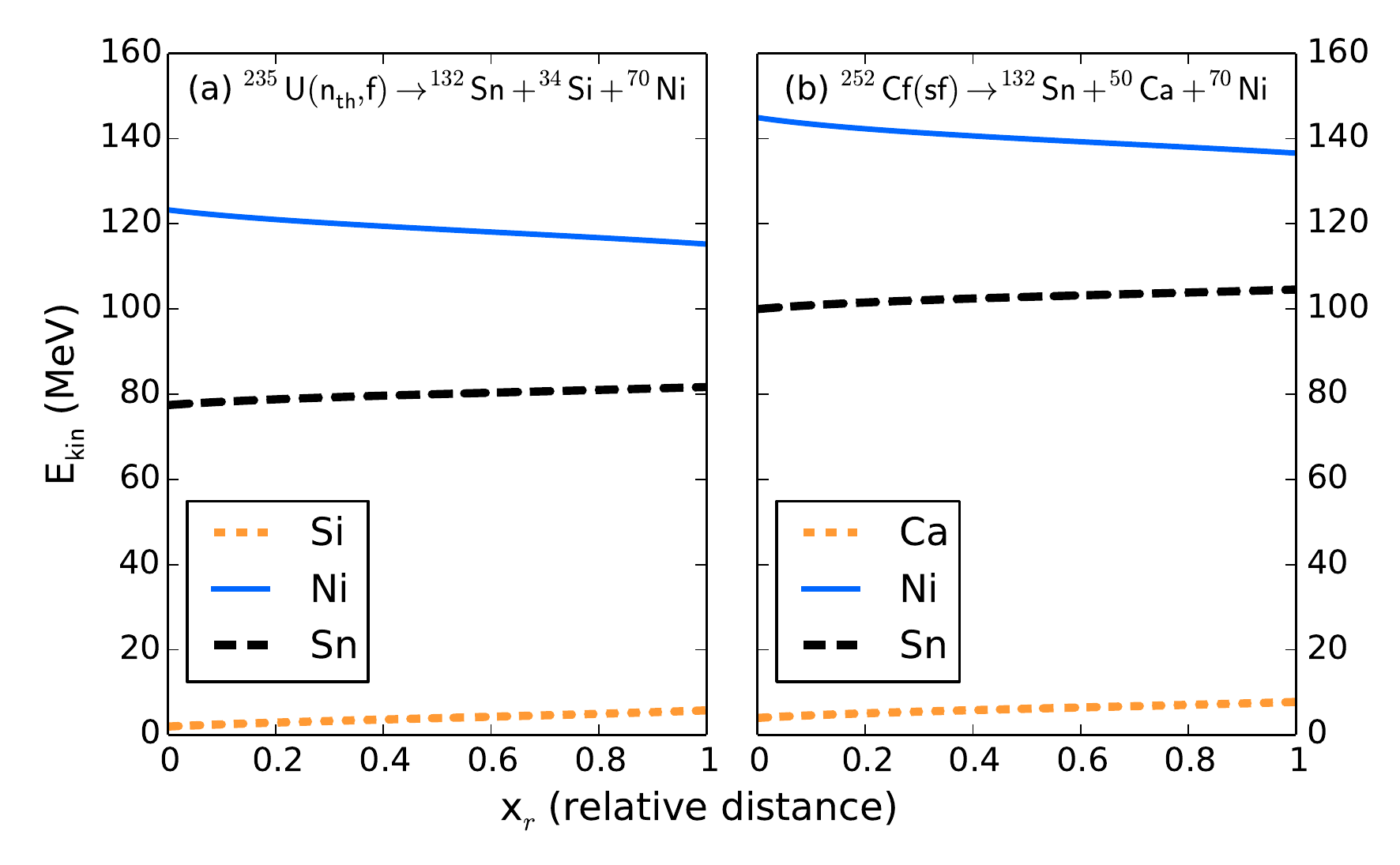}
\caption{\label{fig:simultaneous_single_plot} (Color online) Final fragment kinetic energies versus the relative starting position of the TP, $x_r$, in the true ternary decay model (as described in Sec.~\ref{sec:simultaneous_decay_kinematics}, derivations in App.~\ref{sec:app:ternary}) of (a) ${^{235}\text{U}}(\text{n}_{\text{th}},\text{f})$, and (b) ${^{252}\text{Cf}}(\text{sf})$. When $x_r=0$ and $x_r=1$, the TP starts touching the HF and LF, respectively. Less than $1\%$ of the total energy remains as potential energy, and $\mathit{TXE}=0$~MeV.}
\end{figure}

\begin{figure}[b]
\includegraphics[width=1.0\linewidth]{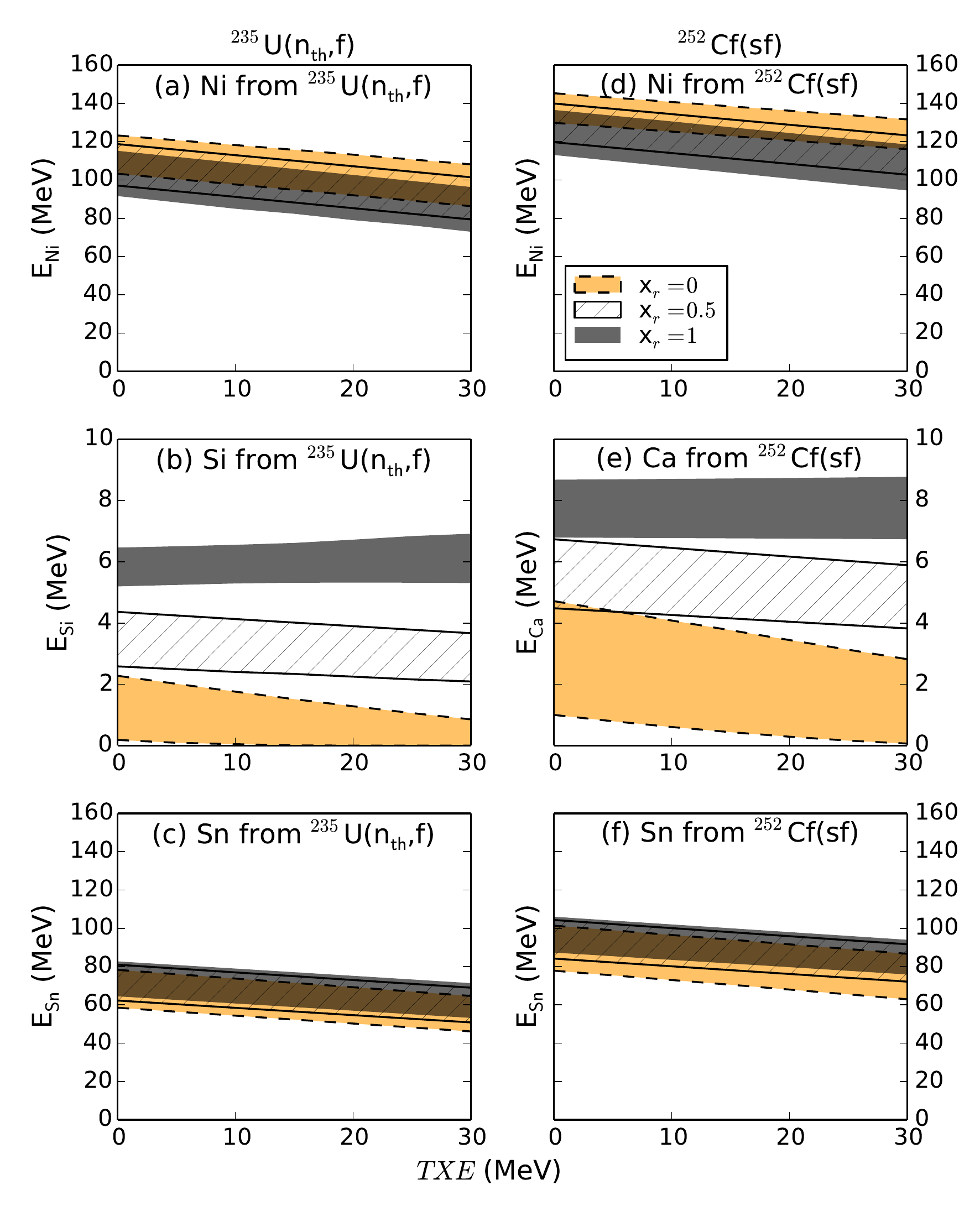}
\caption{\label{fig:simultaneous_areas_all} (Color online) Areas of attainable final kinetic energies of the fission fragments, versus the total excitation energy of all fragments, in the true ternary decays (a--c) ${^{235}\text{U}}(\text{n}_{\text{th}},\text{f}) \rightarrow \text{Sn} + \text{Si} + \text{Ni}$, and (d--f) ${^{252}\text{Cf}}(\text{sf}) \rightarrow \text{Sn} + \text{Ca} + \text{Ni}$. Each figure is element specific, and the different areas indicate the choice of $x_r$, as indicated in the legend in (d). The areas are spanned between the highest and lowest kinetic energies obtained for each choice of $x_r$, by varying the neutron multiplicity as $\nu=0\text{--}4$ and the mass split as $A_{\text{Sn}} = 128\text{--}134$, $A_{\text{Ni}}=68,70,72$, (a--c) $A_{\text{Si}} = 34\text{--}40$, and (d--f) $A_{\text{Ca}} = 42\text{--}56$. Less than $1\%$ of the total energy remains as potential energy.
}
\end{figure}

Using the true ternary decay model described in Sec.~\ref{sec:simultaneous_decay_kinematics} (derivations in App.~\ref{sec:app:ternary}), we parametrize the scission configuration using the fragment mass splits, neutron multiplicity, the relative distance $x_r$ (see Fig.~\ref{fig:simultaneous_decay}) and the total excitation energy of all fragments $\mathit{TXE}=E^{*}_{HF}+E^{*}_{TP}+E^{*}_{LF}$. In Fig.~\ref{fig:simultaneous_single_plot}, the final kinetic energies of the fragments are plotted against $x_r$ for $\mathit{TXE}=0$~MeV.
The highest kinetic energy of the LF is achieved if the TP is formed touching the HF ($x_r=0$), since the TP in this case transfers momentum from the HF to the LF. The central fragment ends up with almost no kinetic energy, since it is confined between the Coulomb forces of the outer fragments. In Fig.~\ref{fig:simultaneous_areas_all}, all final fragment kinetic energies are plotted versus $\mathit{TXE}$.
We represent the kinetic energy range for each fragment with an area, spanned by the highest and lowest kinetic energies obtained. The three different areas represent three different choices of $x_r$, as shown in the legend in Fig.~\ref{fig:simultaneous_areas_all} (d). The ranges of the varied mass splits and neutron multiplicies are given in the caption.

Here, results are presented up to $\mathit{TXE}=30$~MeV. This is higher than the average $\mathit{TXE}$ in alpha accompanied ternary fission. Note that the average $\mathit{TXE}$ also decreases rapidly with increased ternary particle size \cite{bib:mutterer04}. For true ternary fission, any significant $\mathit{TXE}>0$~MeV is therefore not expected.

Our results show that the final kinetic energies generally decrease for increased $\mathit{TXE}$, since there is less energy that can be converted into kinetic energy. The only exception is the kinetic energy of the TP, which increases with increased $\mathit{TXE}$ if it is formed touching the LF ($x_r = 1.0$ in Figs.~\ref{fig:simultaneous_areas_all} (c) and (f)). This is due to the back-scattering dynamics of the ternary particle against the heavy fragment. This dynamics depends on the distance between the light and heavy fragment (which increases with $\mathit{TXE}$), and the ternary particle position between them.

\subsection{\label{sec:collinearity_stability_results}Collinear stability results}
\begin{figure}[b]
\includegraphics[width=1.0\linewidth]{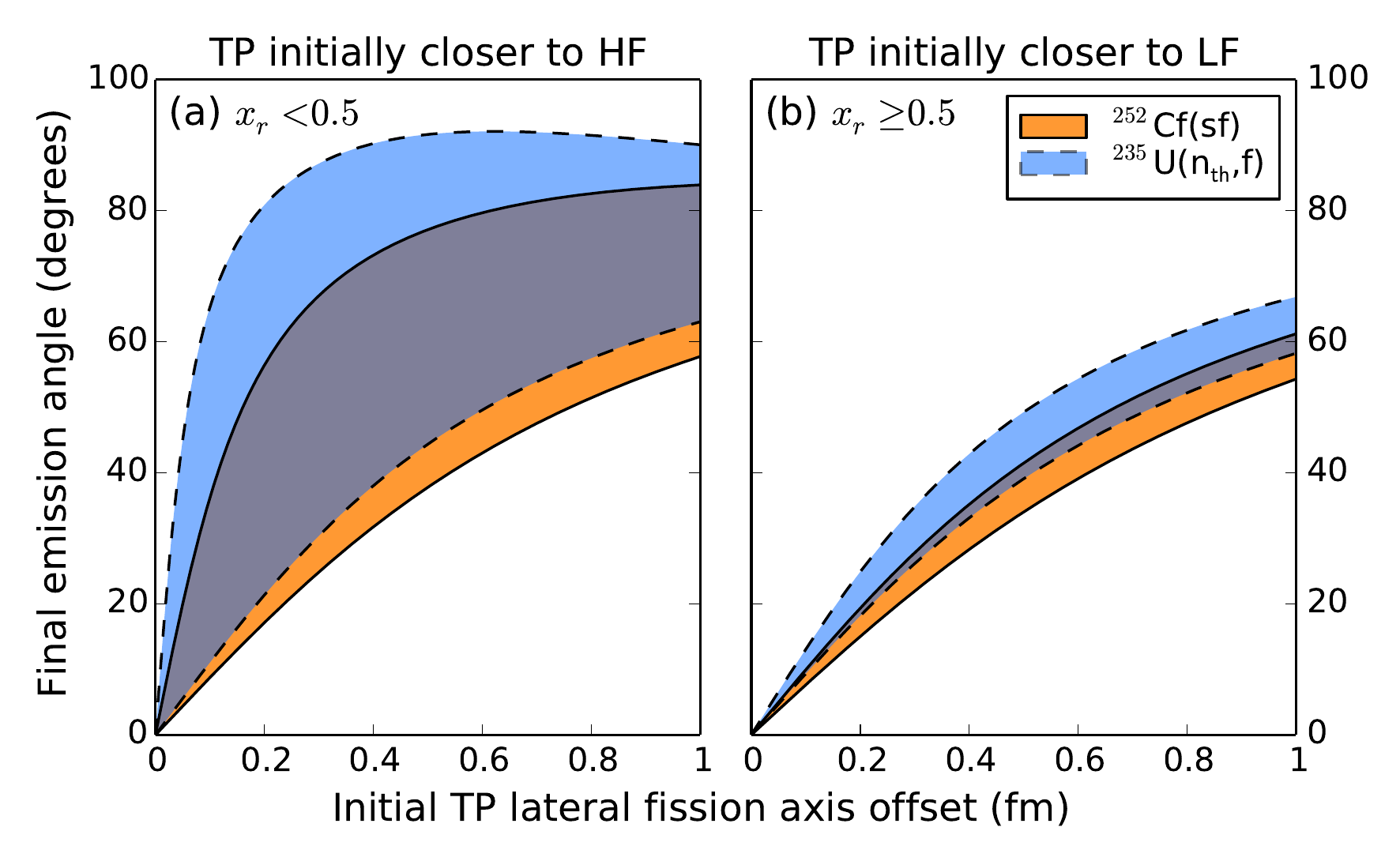}
\caption{\label{fig:angle_vs_y} (Color online) Final emission angle between the light fragment and the ternary particle, versus the initial lateral fission axis offset $y$ of the ternary particle, when the ternary particle is initially closer to (a) the heavy fragment ($x_r < 0.5$), and (b) the light fragment ($x_r \geq 0.5$), respectively. The decays are ${^{235}\text{U}}(\text{n}_{\text{th}},\text{f}) \rightarrow {^{132}\text{Sn}} + {^{34}\text{Si}} + {^{70}\text{Ni}}$ and ${^{252}\text{Cf}}(\text{sf}) \rightarrow {^{132}\text{Sn}} + {^{48}\text{Ca}} + {^{72}\text{Ni}}$, as indicated by the legend. The plots are generated according to App.~\ref{sec:app:stability}, by varying $x_r=0\text{--}1$, $\mathit{TXE}=0\text{--}30$~MeV, and $y=0\text{--}1$~fm. All initial momenta are set to zero.}
\end{figure}

\begin{figure}[b]
\includegraphics[width=1.0\linewidth]{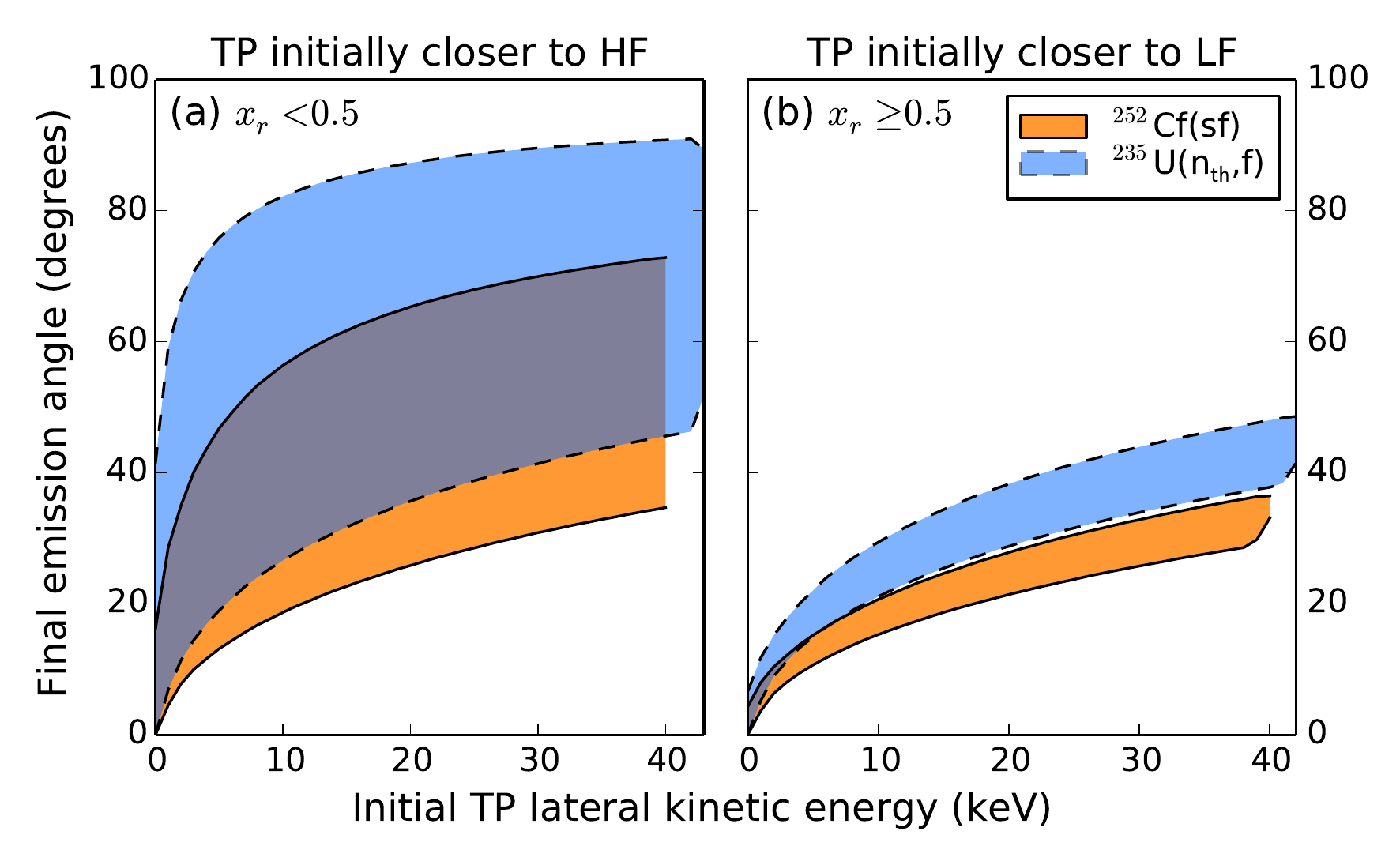}
\caption{\label{fig:angle_vs_py} (Color online) Final emission angle between the light fragment and the ternary particle, versus the initial lateral kinetic energy of the ternary particle, when the ternary particle is initially closer to (a) the heavy fragment ($x_r < 0.5$), and (b) the light fragment ($x_r \geq 0.5$), respectively. The decays are ${^{235}\text{U}}(\text{n}_{\text{th}},\text{f}) \rightarrow {^{132}\text{Sn}} + {^{34}\text{Si}} + {^{70}\text{Ni}}$ and ${^{252}\text{Cf}}(\text{sf}) \rightarrow {^{132}\text{Sn}} + {^{48}\text{Ca}} + {^{72}\text{Ni}}$, as indicated by the legend. The plots are generated according to App.~\ref{sec:app:stability}, by varying $x_r=0\text{--}1$, $\mathit{TXE}=0\text{--}30$~MeV, and $E_{\text{K},0}=0\text{--}0.05$~MeV. The total linear and angular momentum is set to zero, and all particles are initially set perfectly collinearly.}
\end{figure}

Using the Monte-Carlo method described in Sec.~\ref{sec:intrinsic_stability_collinearity} (derivations in App.~\ref{sec:app:stability}), the intrinsic stability of collinearity is examined in the ``true ternary'' decay processes ${^{235}\text{U}}(\text{n}_{\text{th}},\text{f}) \rightarrow {^{132}\text{Sn}} + {^{34}\text{Si}} + {^{70}\text{Ni}}$ and ${^{252}\text{Cf}}(\text{sf}) \rightarrow {^{132}\text{Sn}} + {^{48}\text{Ca}} + {^{72}\text{Ni}}$. The final emission angle between the ternary particle and the light fragment is shown in Fig.~\ref{fig:angle_vs_y} versus the lateral offset of the ternary particle charge-center from the fission axis, denoted $y$, with zero initial momentum, and in Fig.~\ref{fig:angle_vs_py} versus the initial lateral kinetic energy of the ternary particle, i.e. initial momentum $p_y$, when all fragments are formed perfectly collinearly, with zero total linear and angular momentum.
In these figures, each area is spanned by the smallest to largest angles obtained, from more than $10^6$ Monte-Carlo simulations, sampling three different parameters uniformly with $\sim100$ different values each. The first parameter is the ternary particle position between the heavy and light fragment, denoted $x_r$, which is varied from touching the HF ($x_r=0$), to touching the LF ($x_r=1$). The results are shown separately when the ternary particle is formed closer to (a) the HF, and (b) the LF. The second parameter is the sum of the excitation energies of the fragments, denoted $\mathit{TXE} = {E_{HF}^{*}} + {E_{LF}^{*}} + {E_{TP}^{*}}$, which is varied between $0$ and $30$~MeV. The third parameter is the perturbation, being either the lateral ternary particle position ($y$) or momentum ($p_y$). For the position-based perturbation, $y$ is varied between $0$ and $1$~fm, which is much smaller than the TP radius of $r_{TP} = r_{0}\sqrt[3]{A_{TP}} > 4$~fm (where $r_0 \approx 1.25$~fm). The momentum-based perturbation $p_y$ is set by varying the lateral pre-scission kinetic energy of the ternary particle. Note that the total linear and angular momentum can be broken during the time specified by the uncertainty principle. Monte-Carlo simulations were run both conserving and not conserving these momenta, and the results were on the same order of magnitude. Results are shown here (Fig.~\ref{fig:angle_vs_py}) for the former case, with a total lateral pre-scission kinetic energy of the system, denoted $E_{K,0}$, between $0$ and $0.05$~MeV. Note that this initial kinetic energy is very small compared to usual energies in fission. In Fig.~\ref{fig:angle_vs_py}, the initial kinetic energy of the TP extends higher for the lighter ${^{34}\text{Si}}$ than for the heavier ${^{48}\text{Ca}}$, due to momentum conservation.

The calculated angles are expected to hold even for more realistic trajectory calculations, because, as discussed by \citet{bib:carjan80} for the case of alpha-accompanied fission, explicit inclusion of deformations and attractive nuclear interactions leads only to a small ($<10\%$) modification of the resulting angles. For heavier ternary particles than alphas, experiments have shown \cite{bib:grachev80,bib:grachev88,bib:schall88} that with rising nuclear charge of the ternary particle, its angular distribution gets narrower and more perpendicular to the fission axis.

According to our results, the requirement of a collinear emission angle\footnote{Recall that CCT refers to collinear fission events with a relative angle between fragment emission directions of $180\pm2^{\circ}$ \cite{bib:pyatkov10}.} imposes that the charge-center can deviate no more than $y_{\text{TP}}^{\text{max}} \approx 0.02$~fm from the fission axis, or that the lateral initial kinetic energy of the ternary particle can be no larger than $E_{\text{TP}}^{\text{max}} \approx 10^{-4}$~MeV. The corresponding momentum is calculated from
\begin{equation}
\label{eq:stability:results:momentum}
\frac{p_{\text{TP}}^{y,\text{max}}}{\hbar} = \frac{\sqrt{2E_{\text{TP}}^{\text{max}}m_{TP}}}{\hbar},
\end{equation}
with a value of $0.013$ and $0.015$~fm$^{-1}$ for Si and Ca, respectively. From a quantum mechanical perspective, it seems rather unlikely that both the position and momentum could be constrained to such a narrow region. Comparing with the uncertainty principle, we get
\begin{equation}
\label{eq:uncertainty_principle_stability}
\frac{y_{\text{TP}}^{\text{max}}p_{\text{TP}}^{y,\text{max}}}{\hbar} \approx 0.0003 \ll \frac{1}{2} \leq \frac{\Delta{y}\Delta{p_y}}{\hbar},
\end{equation}
for both ${^{34}\text{Si}}$ and ${^{48}\text{Ca}}$.
Only one of the two off-axis dimensions have been included. Introducing the second off-axis dimension reduces the threshold even further, as does including simultaneously a perturbation in position and momentum, or considering only realistic excitation energies close to zero MeV. This implies that collinearity is extremely unstable and improbable in ternary fission due to quantum-mechanical uncertainties.

In addition, we artificially increased the initial separation distance between the outer fragments (the HF and the LF) until the repulsion between them became negligible. These simulations gave similar results, and showed that the repulsion between the TP and the LF alone can break collinearity. Thus, even in a very late sequential decay, collinearity is in question. Assuming a sequential decay, \citet{bib:nasirov16} showed that lateral offsets of the TP up to $2$~fm are expected at scission, and \citet{bib:tashkhodjaev16} confirmed that collinearity indeed can be broken.

\section{\label{sec:discussion}Discussion}
The discussion is separated into three subsections. In the first subsection, we discuss how CCT can be experimentally verified through direct observation, by covering the possible kinetic energy ranges that we have derived. In the second subsection, we discuss the CCT interpretation, and highlight some of its features which are in contrast to previous experiments. In the third subsection, we discuss the intrinsic stability of collinearity, and its consequence for the angular distribution in ternary fission.

\subsection{\label{sec:results_discussion}Verification by direct observation}
Regardless of any decay model a possible CCT could occur through, the kinetic energies must conform to energy and momentum conservation. As apparent in the FOBOS measurements \cite{bib:pyatkov10,bib:pyatkov10_3,bib:pyatkov12}, two fragments will be emitted in opposite directions, with enough energy to be detected by spectrometers. This sets a constraint on the possible kinetic energies of the outer fragments, and consequently leaves little room for considerably changing the kinetic energy distribution of a potential third fragment. Therefore, an experiment aiming at direct observation of CCT with a spectrometer only needs to cover a wide enough energy range to ensure a model-independent verification.

To aid such experiments, we have presented the kinetic energy ranges allowed by energy and momentum conservation, for the collinear emission of three fragments in the processes ${^{235}\text{U}}(\text{n}_{\text{th}},\text{f}) \rightarrow {\text{Sn}} + {\text{Si}} + {\text{Ni}} + \nu{\cdot}\text{n}$, and ${^{252}\text{Cf}}(\text{sf}) \rightarrow {\text{Sn}} + {\text{Ca}} + {\text{Ni}} + \nu{\cdot}\text{n}$. Our derivations allow easy extension to any fissioning system. We have focused on the late sequential and the true ternary decay models, since they represent the extremes of the kinetic energy distribution (illustrated in Figs.~\ref{fig:sequential_areas_all} and \ref{fig:simultaneous_areas_all}, respectively). These kinetic energy ranges obviously extend into physically unreasonable parameter values, and therefore give a safe upper limit for experiments. Therefore, an experiment that can cover the kinetic energies between these extremes can directly verify CCT, and if any events are found, determine the corresponding decay model by comparing the measured kinetic energies to our results. 

These results indicate that it would be relatively easy to use a fission fragment spectrometer to verify CCT through direct observation, by looking for the light fragment, Ni, and in certain configurations, also the ternary particles Si and Ca. Detection of the heavy fragment Sn would be challenging given the similar kinetic energies to binary fission, but more importantly, due to the high background of the latter. At the reported CCT yields of $0.5\%$ \cite{bib:pyatkov10,bib:pyatkov10_3,bib:pyatkov12}, the fragment Ni would be easily discernible against the far-asymmetric fission yield of roughly $10^{-8}$ \cite{bib:kellett09,bib:endf-349}, regardless of kinetic energies. If the ternary particle was formed on the outside, our results indicate that it would have a kinetic energy which is easily separated in a spectrometer, as there is no background of such fragments from binary fission. Several potential energy surface calculations \cite{bib:vijayaraghavan15,bib:oertzen15,bib:tashkhodjaev15} have shown, however, that it would be energetically favorable if the ternary particle was in the center position. In this configuration, all three models show that the kinetic energy of the ternary particle would be close to zero MeV. At such low kinetic energies, the ternary particle would barely be able to leave the target, traverse detector windows and dead layers to deposit enough energy for a clear signal. Hence, it would be very easy to discriminate the ternary particle from the light fragment in a correctly designed experiment, due to the highly differing energies and velocities. Both a symmetric and an asymmetric double-armed detector setup would therefore show a clear missing-mass signature in \textit{both} detectors in a mass-versus-mass spectrum. These results raises the question of why the FOBOS experiments \cite{bib:pyatkov10,bib:pyatkov10_3,bib:pyatkov12} did not see a missing-mass signature in both detector arms, or why \citet{bib:kravtsov99} saw no missing-mass signature from CCT.

\subsection{\label{sec:cct_model_discussion}On the CCT interpretation}
To this date, almost all theoretical research that studies yields and probabilities of CCT, have been based on macroscopic potential-energy surface calculations with few dimensions, see for example Refs.~\cite{bib:vijayaraghavan15,bib:oertzen15,bib:tashkhodjaev15}. In this context it is important to remember a fundamental caveat for all fission models, that was concisely discussed by \citet{bib:moller01}:

\textit{``In the past, fission properties have often been correlated with models of the binding energy of separated fission fragments, and of valleys inside the point of contact. However, the valleys by themselves do not determine the final state of a fissioning nucleus. Final states corresponding to three or more fragments are in many cases energetically more favoured than are states of two final fission fragments. In these cases the nucleus nonetheless divides into only two fragments. This occurs because the barrier between the ground state and the binary fission valley favours such divisions and a ridge separates the binary from the ternary valley, although dynamical effects may also affect the division.''}

Therefore, a full theoretical treatment determining the probability or yield of CCT would require a macroscopic-microscopic or microscopic description with an adequate number of degrees of freedom. State-of-the-art macroscopic-microscopic calculations of potential-energy surfaces in binary fission are based on a 5-dimensional parameterization including elongation, mass asymmetry, neck radius and deformation of the left and the right fragments \cite{bib:moller01,bib:randrup11,bib:ichikawa12,bib:moller15}. To find the fission valleys in this 5-dimensional potential-energy surface, typically 5 million grid points are calculated \cite{bib:moller01}. An extension to CCT with similar details would require in addition five more parameters, namely the mass, deformation and longitudinal position of the ternary fragment, its lateral offset from the fission axis and the second neck radius. Such 10-dimensional potential-energy surface calculations have never been reported and will probably remain beyond computational limitations for some time. The microscopic description by Density Functional Theory (DFT) method also explores a multidimensional deformation landscape by constraining the collective degrees of freedom to the lowest multipole orders (elongation, reflection-asymmetry, necking and triaxiality) \cite{bib:staszczak09}. Again such a description is already very computing intensive for binary fission and a direct extension covering all degrees of freedom required for ternary fission cannot be envisaged at the present.

The original reports by the FOBOS collaboration \cite{bib:pyatkov10,bib:pyatkov10_3,bib:pyatkov12} and the previously mentioned theoretical studies of CCT favor a sequential \cite{bib:vijayaraghavan12} or an ``almost sequential'' \cite{bib:tashkhodjaev15} decay model as an interpretation of the FOBOS experiments. In such a decay, it is not enough to know if the final result has a high Q-value or a low potential-energy surface. The intermediate steps also have to be studied in detail, as they might be forbidden or improbable. The latter was shown to be the case for the masses proposed by FOBOS, as illustrated in our results by the very low, at times even negative, Q-value, of the intermediate fragment. Even if this is compensated for by excitation energy, it does not mean that the fragment will always fission. It also has to overcome or tunnel through the fission barrier, and the probability to fission depends on the barrier penetrability. The fissility of the intermediate fragment ($Z^2/A < 17$ for ${^{104}\text{Mo}}$ and $Z^2/A < 20$ for ${^{120}\text{Cd}}$) is much lower than that of known fissioning systems. The fission barrier can be estimated from the generalized liquid drop model \cite{bib:royer98} to be $44$~MeV for both $^{104}\text{Mo}$ and $^{120}\text{Cd}$. It is known that this model overestimates the real fission barriers \cite{bib:moller95} by up to $10\text{--}12$~MeV.

These fission barriers highlight an overlooked contradiction in the previous theoretical studies on CCT. In a sequential or an almost sequential decay, the intermediate fragment must somehow fission. The low Q-values and the high fission barrier together signify that fission of the intermediate fragment is unlikely. This in turn implies that the yield of binary fission, in which the intermediate fragments Mo and Cd are formed but do not decay, must be significantly higher than the reported yield of CCT of $0.5\%$. No such overabundance of the respective intermediate fragment has been observed. Thus the intermediate Mo and Cd would need to fission with a very high probability, which is in contradiction with the high fission barrier and the low Q-values. To have any probability for fission, a considerable excitation energy would be required. As the excitation energy increases, however, the results have shown that the kinetic energy of the ternary particle approaches zero MeV, leading to the case where CCT should have been found in previous experiments, described in the previous section (Sec.~\ref{sec:results_discussion}). More importantly, at high excitation energies, other modes of de-excitation become competitive, most importantly neutron emission, reducing the likelihood of fission. It is evident that neutron emission in the first fission of the sequential model takes away excitation energy from the system that is essential to enable the second fission step. Therefore it is surprising that FOBOS concluded from their Experiment 3 \cite{bib:pyatkov12} on the isotropic emission of about four neutrons in coincidence with CCT events. To be isotropic these neutrons must be emitted at the first scission, not after pre-acceleration of the IF. The excitation energy of the IF would be correspondingly reduced by about $30$~MeV, i.e. effectively preventing any second fission. Clearly this experimental indication cannot be reconciled with any sequential or near-sequential model.

\subsection{\label{sec:instability_of_collinearity}Intrinsic stability of collinearity}
The discussion has so far highlighted possible short-comings with the currently favored CCT models (the sequential decay models). This inspired us to study a ``true ternary'' decay model. It was shown, however, in Sec.~\ref{sec:collinearity_stability_results} that collinearity is extremely unstable and improbable in ternary fission, since it occurs at an unstable equilibrium point. Thus, it was found that a sufficient perturbation in the ternary particle position or momentum perpendicular to the fission axis breaks collinearity. Furthermore, we found that the threshold in position and momentum for a collinear emission is much smaller than the typical variations governed by the uncertainty principle. The latter could therefore be responsible for breaking collinear configurations into equatorial emission in ternary fission. This means that a collinear (prolate) scission configuration does not necessarily imply collinear emission, and that equatorial emission does not necessarily imply a triangular (oblate) scission configuration. In other words, if CCT occurred through ternary fission, it should have been detected previously by triple-coincidence experiments \cite{bib:schall87}. These arguments might hold even for CCT occurring through late sequential decay, given recent results showing that in this model, it is indeed possible that the ternary particle deviates from the fission axis \cite{bib:nasirov16}, and that collinearity is broken \cite{bib:tashkhodjaev16}.

So far, ternary fission experiments have shown that with rising nuclear charge of the ternary particle, its angular distribution gets narrower and more perpendicular to the fission axis \cite{bib:grachev80,bib:grachev88,bib:schall88}. In addition, the collinearity of the heavy and light fragment is lost due to momentum conservation \cite{bib:gonnenwein05}. This raises the question of why only collinear events but not ternary events with larger angles would be observed for the mass splits proposed by the FOBOS collaboration.

\section{\label{sec:conclusions}Conclusions}
Initially, our model calculations aimed to guide an independent validation experiment by predicting the physically possible kinetic energy range for CCT that has to be covered. Eventually the detailed investigation of currently favored CCT models brought to light features that are either unphysical or that would lead to unique experimental observables, some of which are in contrast to previous experiments.

In the most favorable geometrical configuration, when the lightest fragment is formed at the center, the light fragment has so little energy that it would barely leave the target, traverse windows and dead layers to deposit enough energy in a detector arm. Hence, such CCT events should be clearly visible in any $2v2E$ setup as ``missing mass'' events, even without a specific left-right asymmetry as the support grid in the FOBOS setup. Consequently FOBOS should have observed a similar pattern in both arms and other two-arm-spectrometers should have observed CCT too.

All sequential models implicitly assume a very high fission probability of the intermediate fragment. However, in reality the fissility of the intermediate fragment is low and its fission barrier is very high ($> 30$~MeV). As shown in our calculations, excitation energies of the intermediate fragment far above its fission barrier are unphysical, hence other de-excitation modes (neutron and gamma emission) will largely dominate. Therefore, a given yield of CCT would be accompanied by a much higher yield of binary events involving the specific intermediate fragments that did not undergo fission. Such a peculiar pattern of local over-abundance of specific mass splits should have been detected in previous experiments, in particular in $\gamma \gamma \gamma$-spectroscopy with large Ge detector arrays.

The inclusion of any realistic excitations would further destroy collinearity. For example any fragment angular momentum can turn the fission axis of the intermediate fragment with respect to the first fission axis, thus destroying collinearity.

Collinear configurations are intrinsically unstable in three-body systems with pure or dominantly repulsive forces. Thus, also for true ternary fission models where both scissions happen synchronously, collinearity is unstable. Our calculations quantify this fact by deriving the very restricted phase space of scission configurations that could lead to somehow collinear configurations after acceleration. Even under these simplified model assumptions, all but very few of the ``nearly collinear'' scission configurations will after full acceleration result in large angles between the fragments. Despite the overwhelming number of true ternary fission scission configurations leading to large angles between the fragments, so far no plausible mechanism has been proposed to explain why only CCT events would be observed but not ternary events with larger angles. Our stability calculations imply that the latter should be far more frequent, i.e. leading to an excess yield of ternary mass splits that must have been observed by previous experiments.

In conclusion, our investigation of the currently favored CCT models highlights serious discrepancies of model features with experimental observations. Obviously these very simplified models do not include shell effects, collective excitations, angular momenta, etc. However, in our opinion it is not likely that the explicit inclusion of all these effects could cure the discrepancies of the CCT models, namely the intrinsic instability of collinearity in true ternary fission, or the low fission probability of the intermediate fragment in sequential fission. Thus, we have to conclude that today there is no model that could provide a valid explanation of the experimental claims of CCT put forward by the FOBOS collaboration. Therefore, we encourage further critical theoretical studies with more realistic ternary fission models. More importantly, to truly resolve this matter, experimental verifications are required, which is made possible by covering the kinetic energy ranges derived in this paper, for example with high-resolution fission fragment spectrometers.

\begin{acknowledgments}
We thank Christian Forss\'{e}n and H\r{a}kan T. Johansson for fruitful discussions. P.H. is thankful to the members of the subatomic physics group at Chalmers for their support, and to the staff at the Institut Laue-Langevin for their hospitality extended during his stays in Grenoble.
\end{acknowledgments}

\appendix

\section{\label{sec:app:almost_sequential}Sequential and ``almost sequential'' decay model derivations}
\begin{figure}[b]
\includegraphics[width=0.65\linewidth]{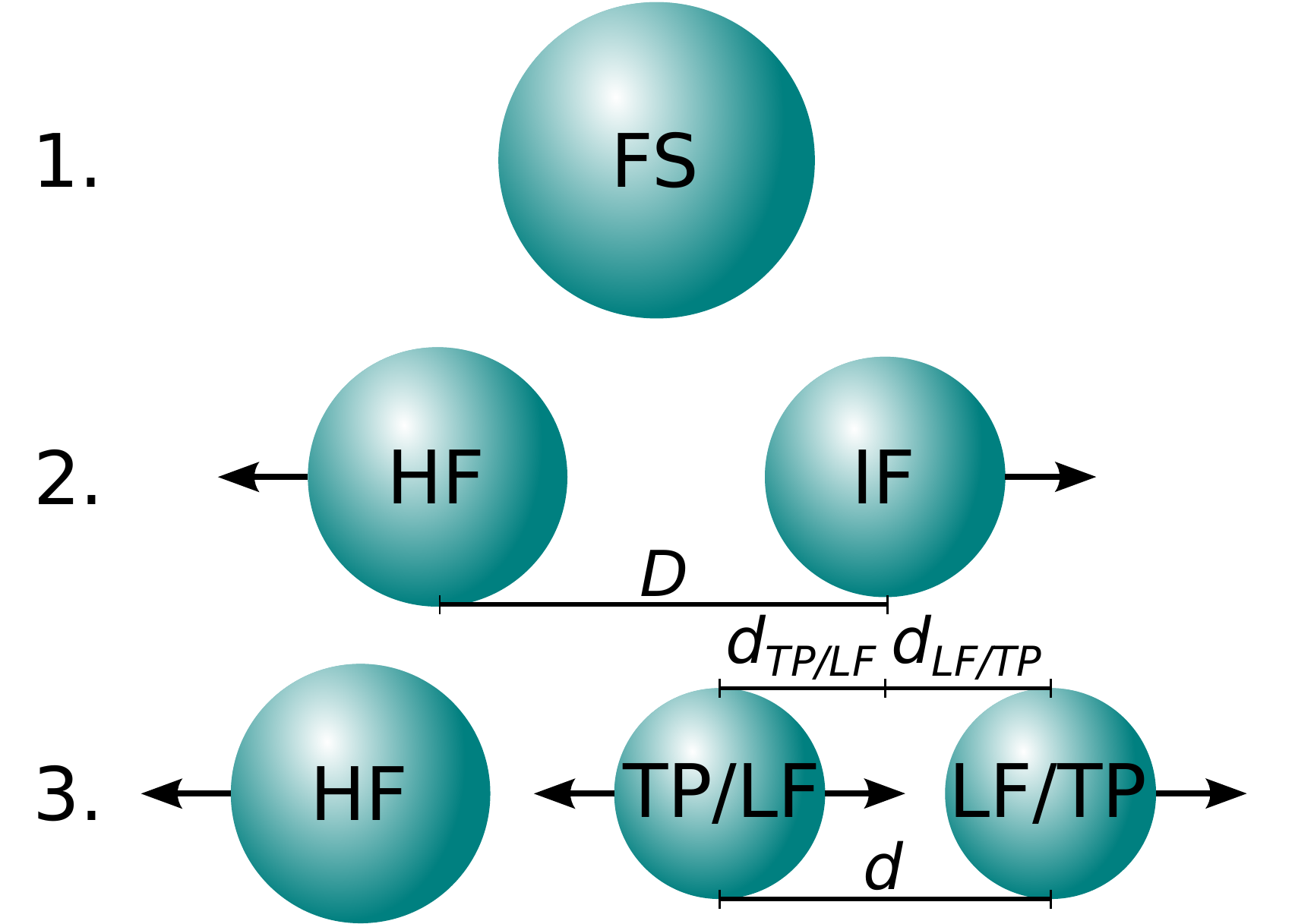}
\caption{\label{fig:app:almost_sequential_decay} (Color online) CCT as a sequential or an ``almost sequential'' decay, as described in Secs.~\ref{sec:sequential_decay_kinematics} and \ref{sec:almost_sequential_decay_kinematics}, respectively. In contrast to the former model, the Coulomb repulsion between the fragments is relevant at all times in the latter model, making the inter-fragment distances relevant to the kinematics. Arrows indicate momentum direction.}
\end{figure}

This appendix describes how the final kinetic energies are solved analytically for CCT fragments from a sequential decay (described in Sec.~\ref{sec:sequential_decay_kinematics}, depicted in Fig.~\ref{fig:app:almost_sequential_decay}), and how the scission-point configuration (initial positions and momenta) of such fragments is constrained in an ``almost sequential'' decay (described in Sec.~\ref{sec:almost_sequential_decay_kinematics}, depicted in Fig.~\ref{fig:app:almost_sequential_decay}). In both cases, this is achieved by energy and momentum conservation.
Both models use the same set of equations, with the difference that in the sequential and almost sequential models, the fragments are fully and partially accelerated before the second scission (the potential energy is zero and non-zero), and the derived kinetic energies are the final and inital (after the second scission), respectively. In the almost sequential model, the final kinetic energies are computed from the scission configuration with semi-classical trajectory calculations, as described in Sec.~\ref{sec:models}. The derivation includes both a repulsive Coulomb potential and an attractive nuclear potential, where a Yukawa-plus-exponential function \cite{bib:krappe79_prl,bib:krappe79_prc} was used for the latter in our calculations. Deformed fragments can be assumed by including the corresponding higher-order moments in these potentials. The kinetic energies carried away by neutrons have not been explicitly included. This is energy which is not available to the acceleration of the final fragments, and can therefore be added to the effective Q-values.

Energy conservation of the first scission gives
\begin{equation}
\label{eq:aseq:q_fs}
Q_{FS} + E^{*}_{FS} = V_1 + E_{HF} + E_{IF} + E^{*}_{HF} + E^{*}_{IF},
\end{equation}
where $V_1$ is the potential, ${E^{*}_{\text{FF}}}$ represents the excitation energy and $E_{\text{FF}}$ the kinetic energy of the respective fragment. The Q-value of the fissioning system is
\begin{equation}
\label{eq:aseq:q1_2}
Q_{FS} = M_{FS} - M_{HF} - M_{IF} - \nu_1 \cdot M_{n},
\end{equation} 
where $M$ are the mass excesses and $\nu_1$ is the neutron multiplicity. The potential is 
\begin{equation}
\label{eq:aseq:pot1}
V_1 = E_{C,1} + E_{N,1}
\end{equation}
where $E_{N,1}$ is the attractive potential, and $E_{C,1}$ the Coulomb potential
\begin{equation}
\label{eq:aseq:ec1}
E_{\text{C},1} = k\frac{e^2Z_{HF}Z_{IF}}{D},
\end{equation}
with $D$ being the center-to-center distance between the HF and the IF, $k$ the Coulomb constant and $e$ the elementary charge. Let ${Q_{FS}^{\text{eff}}}$ denote the ``effective'' Q-value of the first scission (the energy available for acceleration of the HF and the IF)
\begin{equation}
\label{eq:aseq:q1_eff}
{Q_{FS}^{\text{eff}}} = Q_{FS} + {E^{*}_{FS}} - V_1 - {E^{*}_{HF}} - {E^{*}_{IF}}.
\end{equation}
If this term is negative, the corresponding configuration and value of ${E^{*}_{HF}} + {E^{*}_{IF}}$ is energetically forbidden. Let $D_1$ denote the distance at the instant before the scission of the intermediate fragment (note that there is a unique correspondence between this parameter and the time between the two scissions). At this point, the momenta and kinetic energies of the HF and IF are determined from energy and momentum conservation
\begin{eqnarray}
\label{eq:aseq:energy_conservation_1}
E_{HF} + E_{IF} & = & \frac{p^2_{HF}}{2m_{HF}} + \frac{p^2_{IF}}{2m_{IF}}
\\
\label{eq:aseq:momentum_conservation_1}
p_{HF} + p_{IF} & = & 0	,
\end{eqnarray}
with the solutions
\begin{eqnarray}
\label{eq:aseq:HF_mom}
p_{HF} & = & \mp\sqrt{2m_{HF} \cdot {Q_{FS}^{\text{eff}}} \cdot \mu_{IF} }
\\
\label{eq:aseq:IF_mom}
p_{IF} & = & \pm\sqrt{2m_{IF} \cdot {Q_{FS}^{\text{eff}}} \cdot \mu_{HF} }
\\
\label{eq:aseq:HF_kin}
E_{HF} & = & {Q_{FS}^{\text{eff}}} \cdot \mu_{IF}
\\
\label{eq:aseq:IF_kin}
E_{IF} & = & {Q_{FS}^{\text{eff}}} \cdot \mu_{HF},
\end{eqnarray}
where the reduced masses are
\begin{eqnarray}
\label{eq:aseq:reduced_HF}
\mu_{HF} & = & \frac{m_{HF}}{m_{HF}+m_{IF}}\\
\label{eq:aseq:reduced_IF}
\mu_{IF} & = & \frac{m_{IF}}{m_{HF}+m_{IF}}.
\end{eqnarray}
Eqs.~(\ref{eq:aseq:HF_kin}) and (\ref{eq:aseq:IF_kin}) are the final kinetic energies of fully accelerated fragments if $V_1(D_1\to\infty)\to0$. The positive and negative signs of the momenta correspond to the two directions along the fission axis. After the second scission, $D$ denotes the distance between the HF and the mass-center of the TP and the LF. Let $d$ denote the charge-center distance between the TP and the LF
\begin{equation}
\label{eq:aseq:d}
d = d_{TP} + d_{LF},
\end{equation}
where $d_{TP}$ and $d_{LF}$ are the center-of-mass distances to the TP and LF, respectively, related by
\begin{equation}
\label{eq:aseq:com_conservation}
-d_{TP}m_{TP} + d_{LF}m_{LF} = 0.
\end{equation}
Combining Eqs.~(\ref{eq:aseq:com_conservation}) and (\ref{eq:aseq:d}) yields
\begin{eqnarray}
\label{eq:aseq:d_tp_lf_2}
d_{LF} & = & d \cdot \mu_{TP}
\\
\label{eq:aseq:d_tp_lf_1}
d_{TP} & = & d \cdot \mu_{LF},
\end{eqnarray}
where the reduced masses are
\begin{eqnarray}
\label{eq:aseq:reduced_LF}
\mu_{LF} & = & \frac{m_{LF}}{m_{LF}+m_{TP}}\\
\label{eq:aseq:reduced_TP}
\mu_{TP} & = & \frac{m_{TP}}{m_{LF}+m_{TP}}.
\end{eqnarray}
Energy conservation of the second scission gives
\begin{multline}
\label{eq:aseq:q2_1}
Q_{IF} + V_1 + E_{IF} + E_{IF}^{*} =\\V_2 + E_{TP} + E_{LF} + E_{TP}^{*} + E_{LF}^{*},
\end{multline}
and energy conservation of both fission processes give
\begin{equation}
\label{eq:aseq:q_full}
Q_{FS} + E^{*}_{FS} + Q_{IF} = V_2 + \mathit{TKE} + \mathit{TXE}
\end{equation}
where
\begin{eqnarray}
\label{eq:aseq:pot1}
V_2 & = & E_{C,2} + E_{N,2}\\
\label{eq:aseq:TKE}
\mathit{TKE} & = & E_{HF} + E_{LF} + E_{TP}\\
\label{eq:aseq:TXE}
\mathit{TXE} & = & {E^{*}_{HF}} + {E^{*}_{LF}} + {E^{*}_{TP}}.
\end{eqnarray}
The Q-value is given by
\begin{equation}
\label{eq:aseq:q2_2}
Q_{IF} = M_{IF} - M_{LF} - M_{TP} - \nu_2 \cdot M_{n},
\end{equation}
and the Coulomb energy after the second scission is
\begin{eqnarray}
E_{\text{C},2} = k\frac{e^2Z_{TP}Z_{HF}}{D - d\cdot \mu_{LF}} & + & k\frac{e^2Z_{LF}Z_{HF}}{D+d\cdot \mu_{TP}}\nonumber\\
\label{eq:aseq:ec2}
& + & k\frac{e^2Z_{TP}Z_{LF}}{d}.
\end{eqnarray}
Rearranging the equations slightly yields
\begin{multline}
\label{eq:aseq:ekin_tplf}
E_{TP}+E_{LF} =\\Q_{IF} + E^{*}_{IF} + E_{IF} + V_1 - V_2 - E^{*}_{TP} - E^{*}_{LF},
\end{multline}
where from now on $V_1$ is evaluated at $D_1$ and $V_2$ at any arbitrary $D \geq D_1$.
The momenta and kinetic energies of the TP and LF are determined from energy and momentum conservation
\begin{eqnarray}
\label{eq:aseq:ekin_tplf_2}
E_{TP}+E_{LF} & = & \frac{p^{2}_{TP}}{2m_{TP}} + \frac{p^{2}_{LF}}{2m_{LF}}\\
\label{eq:aseq:mom_cons_2}
p_{LF} + p_{TP} & = & p_{IF},
\end{eqnarray}
with the solutions
\begin{eqnarray}
\label{eq:aseq:p_lf}
p_{LF} & = & p_{IF}\cdot\mu_{LF} \pm \sqrt{2m_{LF} \cdot {Q_{IF}^{\text{eff}}} \cdot \mu_{TP}}\\
\label{eq:aseq:p_tp}
p_{TP} & = & p_{IF}\cdot\mu_{TP} \mp \sqrt{2m_{TP} \cdot {Q_{IF}^{\text{eff}}} \cdot \mu_{LF}}.
\\
\label{eq:aseq:LF_kin}
E_{LF} & = & \left(\sqrt{E_{IF} \cdot \tilde{\mu}_{IF} \cdot \mu_{LF}} \pm \sqrt{ {Q_{IF}^{\text{eff}}} \cdot \mu_{TP} }\right)^2\\
\label{eq:aseq:TP_kin}
E_{TP} & = & \left(\sqrt{E_{IF} \cdot \tilde{\mu}_{IF} \cdot \mu_{TP}} \mp \sqrt{ {Q_{IF}^{\text{eff}}} \cdot \mu_{LF} }\right)^2
\end{eqnarray}
where ${Q_{IF}^{\text{eff}}}$ is the effective Q-value of the second fission (the energy available for acceleration of the TP, LF and the HF, if not fully accelerated already)
\begin{equation}
\label{eq:aseq:q2_eff}
{Q_{IF}^{\text{eff}}} = Q_{IF} + V_1 - V_2 + E^{*}_{IF} - E^{*}_{TP} - E^{*}_{LF} - E_{n}
\end{equation}
with the kinetic energy of the emitted neutrons
\begin{equation}
\label{eq:aseq:n_kin}
E_{n} = E_{IF}\frac{\nu_2m_{n}}{m_{LF}+m_{TP}}.
\end{equation}
and the reduced mass $\tilde{\mu}_{IF}$
\begin{equation}
\label{eq:aseq:reduced_IF_2}
\tilde{\mu}_{IF} = \frac{m_{IF}}{m_{LF}+m_{TP}}.
\end{equation}
The positive and negative signs of the momenta represent an increase and decrease in momentum corresponding to the outer and inner positions, respectively. This is because the acceleration of the inner fragment is opposite to the direction that the IF was moving in, leaving it with little to zero kinetic energy. Conversely, the outer fragment is accelerated in the same direction as the IF was moving in, boosting its kinetic energy. The boost is increased as $E^{*}_{IF}$ increases. The kinetic energy of the HF follows the opposite trend, since any excitation energy steals energy from the first scission, leaving less for kinetic energy.

If at the moment of the second scission $V_1(D_1\to\infty)\to0$, and subsequently $V_2(D,d\to\infty)\to0$, then the kinetic energies in Eqs.~(\ref{eq:aseq:HF_kin}), (\ref{eq:aseq:LF_kin}) and (\ref{eq:aseq:TP_kin}) are the final kinetic energies of the HF, LF and TP, respectively. On the other hand, if $D_1$ is finite at the moment of the second scission such that the potential energy of the HF is not negligible, then the scission-point configuration is specified by $D_1$ and $d$ together with the momenta in Eqs.~(\ref{eq:aseq:HF_mom}), (\ref{eq:aseq:p_lf}) and (\ref{eq:aseq:p_tp}). The configuration is uniquely constrained by energy and momentum conservation for given excitation energies, mass splits and neutron multiplicities. The final kinetic energies in turn depend on exactly how $D$ and $d$ are taken to infinity. This is contained in the dynamics but not the kinematics, which is why semi-classical trajectory calculations are used.

\section{\label{sec:app:ternary}True ternary decay model derivations}
\begin{figure}[b]
\includegraphics[width=0.65\linewidth]{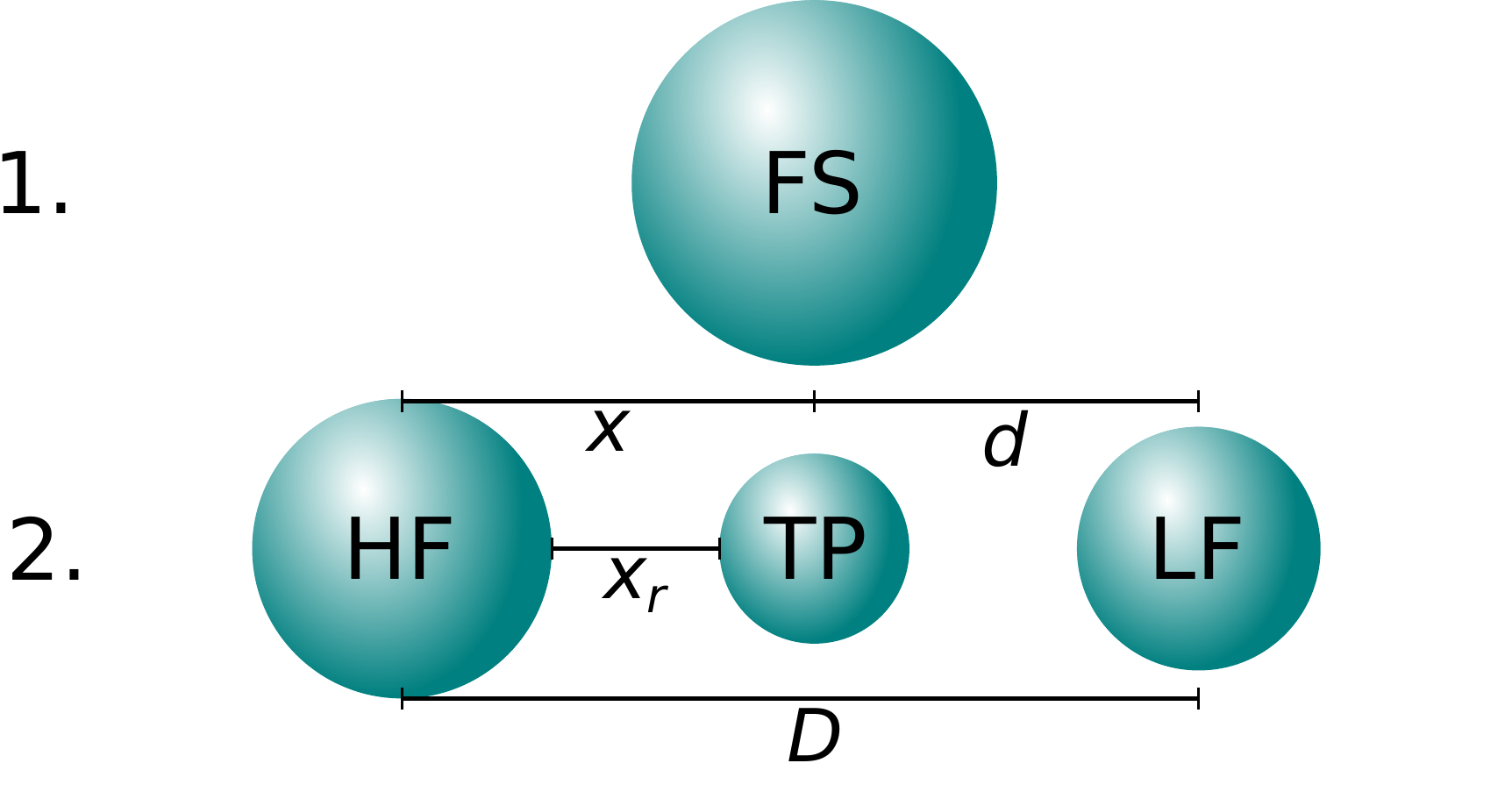}
\caption{\label{fig:app:simultaneous_decay} (Color online) CCT as a true ternary decay, as described in Sec.~\ref{sec:simultaneous_decay_kinematics}. Note that $x_r$ is a relative and dimensionless coordinate, defined between $0$ and $1$, corresponding to the TP ``touching'' the HF and the LF, respectively.}
\end{figure}

This appendix describes how the scission-point configuration is constrained by energy and momentum conservation for given CCT fragments in the true ternary decay model (see Fig.~\ref{fig:app:simultaneous_decay}). The final kinetic energies are computed from this scission configuration with semi-classical trajectory calculations, as described in Sec.~\ref{sec:models}.

Energy conservation of the fission gives
\begin{equation}
\label{eq:q_simultaneous}
Q_{FS} + {E^{*}_{FS}} = V + \mathit{TXE} + E_{\text{K},0},
\end{equation}
where $\mathit{TXE} = {E_{HF}^{*}} + {E_{LF}^{*}} + {E_{TP}^{*}}$ is the sum of all fragment excitation energies, $E_{\text{K},0}$ is the total pre-scission kinetic energy, ${E^{*}_{FS}}$ the excitation energy of the fissioning system, and $V$ is the potential energy
\begin{equation}
\label{eq:trueternary:potential}
V = E_{\text{C}} + E_{\text{N}}.
\end{equation}
The Q-value is
\begin{equation}
\label{eq:q2_simultaneous}
Q_{FS} = M_{FS} - M_{HF} - M_{TP} - M_{LF} - \nu \cdot M_{n},
\end{equation}
where $M$ are the mass excesses and $\nu$ is the neutron multiplicity. The Coulomb potential is given by
\begin{equation}
\label{eq:ec_simultaneous}
E_{\text{C}} = k\frac{e^2Z_{TP}Z_{HF}}{x} + k\frac{e^2Z_{TP}Z_{LF}}{D-x} + k\frac{e^2Z_{HF}Z_{LF}}{D},
\end{equation}
where $k$ is the Coulomb constant, $e$ the elementary charge, and the center-to-center distances are $x$ between the TP and HF, $D$ between the HF and LF, and $D-x$ between the TP and LF. $E_{\text{N}}$ is the attractive nuclear potential. Note that we can express $x$ in terms of $D$ and the relative (and dimensionless) coordinate $x_r\in[0,1]$ as
\begin{equation}
\label{eq:x_simultaneous}
x = r_{TP} + r_{HF} + x_r\left(D - 2r_{TP} - r_{LF} - r_{HF}\right),
\end{equation}
where $r_{TP}$, $r_{HF}$ and $r_{LF}$ denote the radii of the respective fragments. To obtain the scission configuration, Eqs.~(\ref{eq:q2_simultaneous})--(\ref{eq:x_simultaneous}) are substituted into Eq.~(\ref{eq:q_simultaneous}) and solved with respect to $D$. There are multiple solutions corresponding to all possible fragment arrangements, only one of which corresponds to the arrangement in Fig.~\ref{fig:simultaneous_decay}. Choosing this $D$ fully determines the scission-point configuration through the above equations, for a given fragment mass split, neutron multiplicity and pre-scission kinetic energy, with the parameters $x_r$ and $\mathit{TXE}$.

\section{\label{sec:app:stability}Intrinsic stability of collinearity derivations}
\begin{figure}[b]
\includegraphics[width=0.65\linewidth]{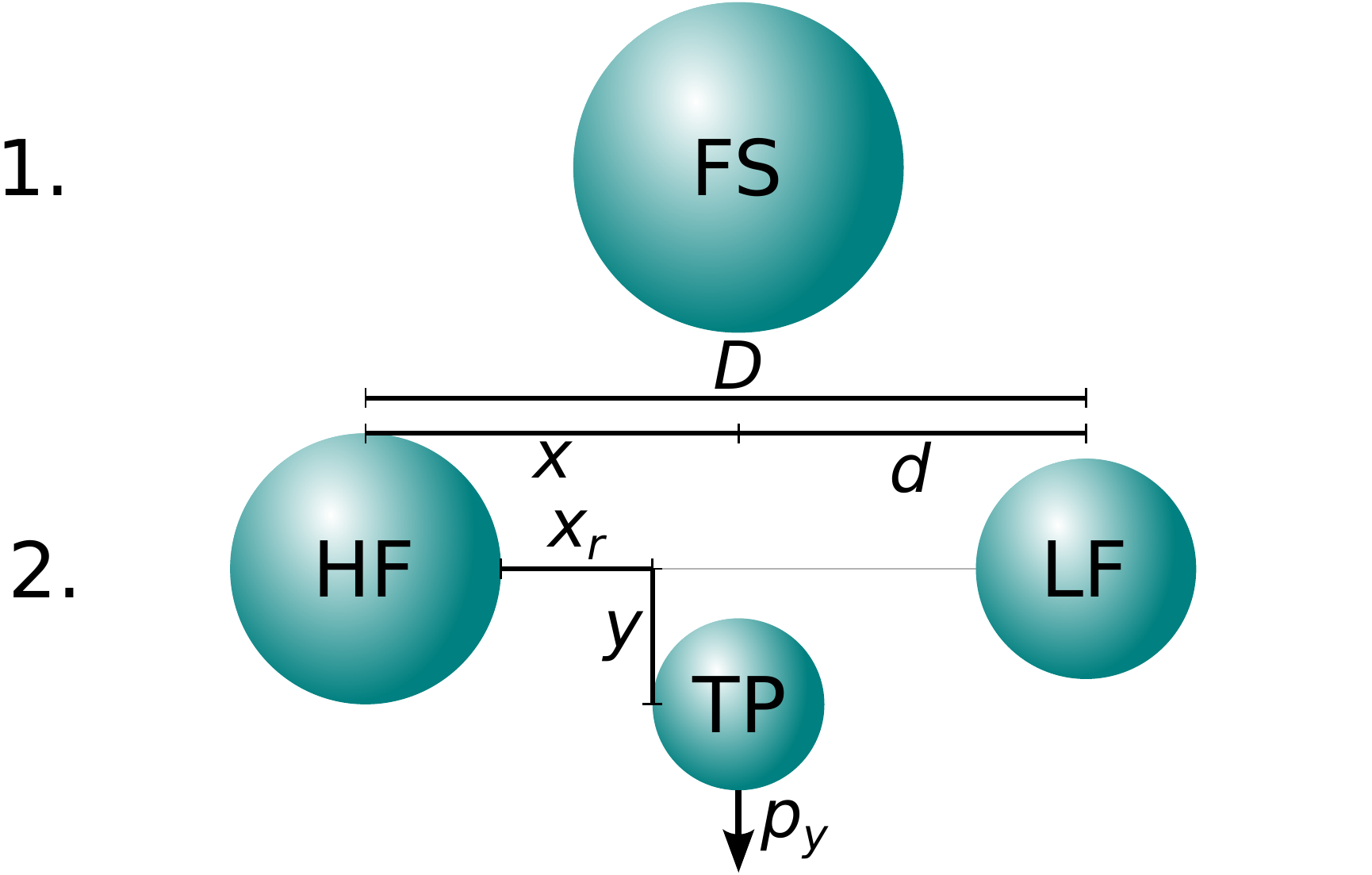}
\caption{\label{fig:app:stability_simultaneous_decay} (Color online) CCT as a true ternary decay with a perturbation in the ternary particle position or momentum, as described in Sec.~\ref{sec:intrinsic_stability_collinearity}. Note that $x_r$ is a relative and dimensionless coordinate, defined between $0$ and $1$, corresponding to the TP ``touching'' the HF and the LF, respectively.}
\end{figure}

Based on the equations of App.~\ref{sec:app:ternary}, this appendix describes how the scission-point configuration is constrained by energy and momentum conservation for given CCT fragments in the true ternary decay model, with a perturbation in the ternary particle position or momentum (see Fig.~\ref{fig:app:stability_simultaneous_decay}). The final kinetic energies are computed from this scission configuration with semi-classical trajectory calculations, as described in Sec.~\ref{sec:models}.
For the momentum-based perturbation, the ternary particle offset from the fission axis is set to $y=0$, and the total pre-scission kinetic energy
\begin{equation}
\label{eq:stability:ekin}
E_{\text{K},0} = \frac{p^2_{TP}}{2m_{TP}} + \frac{p^2_{HF}}{2m_{HF}} + \frac{p^2_{LF}}{2m_{LF}}
\end{equation}
in Eq.~(\ref{eq:q_simultaneous}) is sampled as $E_{\text{K},0} \in [0,{E_{\text{K},0}^{\text{max}}}]$, where ${E_{\text{K},0}^{\text{max}}}$ is a limit to be set. The relative ternary particle position $x_r$ is sampled as $x_r \in [0, 1]$, and the sum of the excitation energies of the fragments as $\mathit{TXE} \in [0, \mathit{TXE}_{\text{max}}]$, where $\mathit{TXE}_{\text{max}}$ is a limit to be set. With these parameters given, all the inter-particle distances are fully determined as described in the true ternary decay model (App.~\ref{sec:app:ternary}). It is now described how to determine the initial momenta. Let the origin of the lab frame be the charge-center of the ternary particle. Let the $x$-axis be the fission axis, and the initial momenta be along the $y$-axis. The coordinate of the center-of-mass in this lab frame is
\begin{equation}
\label{eq:stability:com}
R_{\text{CM}} = \frac{-x \cdot m_{HF} + (D-x) \cdot m_{LF}}{m_{TP} + m_{HF} + m_{LF}},
\end{equation}
The particle coordinates in the center-of-mass frame are
\begin{eqnarray}
R_{TP} & = & -R_{\text{CM}}\label{eq:stability:com_tp},\\
R_{HF} & = & R_{TP} - x\label{eq:stability:com_hf},\\
R_{LF} & = & R_{TP} + (D-x)\label{eq:stability:com_lf}.
\end{eqnarray}
Conservation of linear and angular momenta yield
\begin{eqnarray}
p_{TP} + p_{HF} + p_{LF} & = & 0\label{eq:stability:linear_mom}\\
R_{TP} \cdot p_{TP} + R_{HF} \cdot p_{HF} + R_{LF} \cdot p_{LF} & = & 0,\label{eq:stability:angular_mom}
\end{eqnarray}
respectively, which gives the relations
\begin{eqnarray}
p_{HF} = \eta_{1}p_{TP}\label{eq:stability:phf}\\
p_{LF} = \eta_{2}p_{TP}\label{eq:stability:plf},
\end{eqnarray}
where
\begin{eqnarray}
\eta_{1} = \frac{R_{LF} - R_{TP}}{R_{HF} - R_{LF}}\label{eq:stability:eta1}\\
\eta_{2} = \frac{R_{HF} - R_{TP}}{R_{LF} - R_{HF}}\label{eq:stability:eta2}.
\end{eqnarray}
Inserting these relations into Eq.~(\ref{eq:stability:ekin}), the initial momentum of the ternary particle is found to be
\begin{equation}
\label{eq:stability:ptp}
p_{TP} = \sqrt{2E_{\text{K},0}\left(\frac{1}{m_{TP}} + \frac{\eta_1}{m_{HF}} + \frac{\eta_2}{m_{LF}}\right)^{-1}},
\end{equation}
where the TP momentum has been chosen to be along the positive $y$-direction.

For the position-based perturbation, the pre-scission kinetic energy is set to zero, and the ternary particle is offset from the fission axis with a distance $y \in [0, y_{max}]$, where $y_{max}$ is a limit to be set. This allows for a tighter scission configuration, giving a higher off-axis repulsion (which more easily breaks collinearity), but this effect was ignored in this paper.

\providecommand{\noopsort}[1]{}\providecommand{\singleletter}[1]{#1}%

\end{document}